\documentclass[pdflatex,sn-mathphys-ay]{sn-jnl}

\usepackage{graphicx}%
\usepackage{multirow}%
\usepackage{amsmath,amssymb,amsfonts}%
\usepackage{amsthm}%
\usepackage{mathrsfs}%
\usepackage[title]{appendix}%
\usepackage{xcolor}%
\usepackage{textcomp}%
\usepackage{manyfoot}%
\usepackage{booktabs}%
\usepackage{algorithm}%
\usepackage{algorithmicx}%
\usepackage{algpseudocode}%
\usepackage{listings}%

\usepackage{subfig}
\usepackage{grffile}

\usepackage{relsize}

\theoremstyle{thmstyleone}%
\theoremstyle{thmstyletwo}%

\theoremstyle{thmstylethree}%

\usepackage{xcolor}
\colorlet{darkgreen}{green!40!black}
\colorlet{suscol}{darkgreen}
\colorlet{infonecol}{red}
\colorlet{inftwocol}{yellow!80!black}
\colorlet{newinftwocol}{purple}
\colorlet{rescol}{darkgray}

\colorlet{drivecol}{brown!50!black}
\colorlet{parkcol}{blue}

\colorlet{eqbg}{white}

\newcommand{\gO}[1][]{\textcolor{darkgreen}{O$_{#1}$}}

\newcommand{\rC}[1][]{\textcolor{red}{C$_{#1}$}}

\newcommand{\dyn}{\mathcal{D}}

\newcommand{\nO}{n_{\text{\gO}}}
\newcommand{\nC}{n_{\text{\rC}}}

\newcommand{\nE}{n_{\text{E}}}
\newcommand{\nV}{n_{\text{V}}}

\newcommand{\tO}[1][]{t_{\text{\gO}}^{#1}}

\newcommand{\tC}[1][]{t_{\text{\rC}}^{#1}}

\newcommand{\fO}{f_\text{\gO}}
\newcommand{\fC}{f_\text{\rC}}
\newcommand{\fOC}{f_\text{\gO\rC}}
\newcommand{\fCO}{f_\text{\rC\gO}}

\newcommand{\To}{T_{\text{\gO}}}

\newcommand{\Tc}{T_{\text{\rC}}}
\newcommand{\Dc}{D_{\text{\rC}}}

\newcommand{\lamC}[1][]{ {\lambda^{#1}_{\text{\rC}} } }
\newcommand{\lamO}[1][]{\lambda^{#1}_{\text{\gO}}}

\newcommand{\prob}{\mathbb{P}}

\newcommand{\ipthree}{IP$_3$}
\newcommand{\ipr}{\ipthree R}

\colorlet{actcol}{brown}
\colorlet{inactcol}{blue}

\DeclareMathOperator{\Det}{det}
\DeclareMathOperator{\Tr}{Tr}

\usepackage{tikz}
\usetikzlibrary{arrows,calc,fit, shapes,shapes.symbols,positioning, decorations.pathmorphing,decorations.markings}

\begin{document}

\title[Ion channels \& identifiability]{Modelling ion channels with a
  view towards identifiability\footnote{Dedicated to Edmund~J. Crampin
  (1973-2021)}}


\author*[1]{\fnm{Ivo} \sur{Siekmann}}\email{i.siekmann@ljmu.ac.uk}



\affil*[1]{\orgdiv{School of Computer Science and Mathematics (CSM)},
  \orgname{Liverpool John Moores University (LJMU)},
  \orgaddress{\street{Byrom Way}, \city{Liverpool}, \postcode{L3 3AF},
    \state{Merseyside}, \country{United Kingdom}}}




\abstract{Aggregated Markov models provide a flexible framework for
  stochastic dynamics that develops on multiple timescales. For
  example, Markov models for ion channels often consist of multiple
  open and closed state to account for ``slow'' and ``fast'' openings
  and closings of the channel. The approach is a popular tool in the
  construction of mechanistic models of ion channels---instead of
  viewing model states as generators of sojourn times of a certain
  characteristic length, each individual model state is interpreted as
  a representation of a distinct biophysical state.  We will review
  the properties of aggregated Markov models and discuss the
  implications for mechanistic modelling. First, we show how the
  aggregated Markov models with a given number of states can be
  calculated using P\'{o}lya enumeration. However, models with~$\nO $
  open and~$\nC$ closed states that exceed the maximum
  number~$2\nO \nC$ of parameters are non-identifiable. We will
  present two derivations for this classical result and investigate
  non-identifiability further via a detailed analysis of the
  non-identifiable fully connected three-state model. Finally, we will
  discuss the implications of non-identifiability for mechanistic
  modelling of ion channels. We will argue that instead of designing
  models based on assumed transitions between distinct biophysical
  states which are modulated by ligand binding, it is preferable to
  build models based on additional sources of data that give more
  direct insight into the dynamics of conformational changes.
}




\maketitle

\section{Introduction}
\label{sec:intro}

Ion channels belong to the simplest components of living cells---in
membranes that are usually impermeable to ions they open tiny pores
which allow ions to enter or leave the cell. Many channels select
which ions can pass through the pore but otherwise, ion channels exert
little influence on the net flux of ions which is determined by a
combination of the concentration gradient and the membrane potential
i.e. the electrochemical gradient. Thus, at the most basic level, ion
channels can be viewed as switches which are able to switch on or off
fluxes of ions that would otherwise be blocked by the cell membrane.

Despite, or maybe due to their simplicity which enables them to be
flexibly integrated in various physiological systems, ion channels are
involved in more or less all important physiological
processes
. 
These can be roughly summarised in three
categories:

\begin{enumerate}
\item Regulating the intracellular concentration of a particular ion.
\item Regulating the charge within a cell i.e. the membrane potential.
\item Signal transduction.
\end{enumerate}



Ion channels belong to a class of cellular building blocks called
membrane proteins. 
Although the function carried out by an ion channel can be 
described as one of a simple switch, the membrane protein which
carries out this task is not at all simple at a biophysical level---to
open a pore in the membrane, the channel protein needs to undergo one
or more rearrangements of its complex three-dimensional
structure---so-called conformational changes.

A fundamental question in ion channel modelling is

\begin{quotation}
``Should the
\emph{biophysics of the channel protein} be reflected in a
mathematical model describing the \emph{dynamics of the ion
  channel}?''  
\end{quotation}

We initially disregard this important question but will return to
consider it after our detailed investigation of non-identifiability,
see the Discussion, sec.~\ref{sec:discussion}.

We describe the dynamics $\dyn$ of an ion channel as an alternating
sequence of open and closed times,
$\dyn = \left( \tO[1], \tC[1], \tO[2], \tC[2], \dots\ \right)$
(Without loss of generality we assume that the first time observed is
an open time but we could equally have started with a closed
time). The times $\tO[1], \tC[1], \dots$ are stochastic---when an ion
channel opens, it is, in principle, not possible to predict, how long
it will stay open.

\begin{figure}[htbp]
  \centering
  \subfloat[Stochastic ion channel dynamics~$\dyn$]{
    \begin{minipage}[b]{0.45\linewidth}
      \centering
    \begin{tikzpicture}[scale=1]
      \draw[|-|, darkgreen, thick]    (0pt,0) --
      node[below]{\scriptsize $\tO[1]$
      } (1cm,0);
      \draw[|-|, red, thick]    (1.05cm,0) --
      node[below]{\scriptsize $\tC[1]$
      } (2.5cm,0);
      
      \draw[|-|, darkgreen, thick]    (2.55cm,0) --
      node[below]{\scriptsize $\tO[2]$
      } (4.0cm,0);
      \draw[|-|, red, thick]    (4.05cm,0) --
      node[below]{\scriptsize $\tC[2]$
      } (5.5cm,0);
      
      \draw[|-|, darkgreen, thick]    (5.55cm,0) --
      node[below]{\scriptsize $\tO[3]$
      } (6.0cm,0);

      \path node at (3.0cm,0.3cm){};
    \end{tikzpicture}
  \end{minipage}
} \;
  \subfloat[Continuous-time two-state Markov model]{%
    \begin{minipage}[b]{0.45\linewidth}
      \centering
      \includegraphics[width=0.2\linewidth]{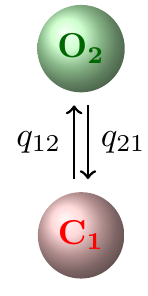}%
      
    \end{minipage}
  }%
  \caption{(a) The dynamics~$\dyn$ of an ion channel is an
    alternating sequence of stochastic open and closed times~$\tO[i]$
    and~$\tC[j]$. (b) Such a sequence can be described by a
    continuous-time Markov model with one closed state~\rC[1] and one
    open state \gO[2]. If the ion channel generates open or closed
    times of multiple characteristic lengths, additional open or
    closed states need to be added.}
  \label{fig:icdyn}
\end{figure}
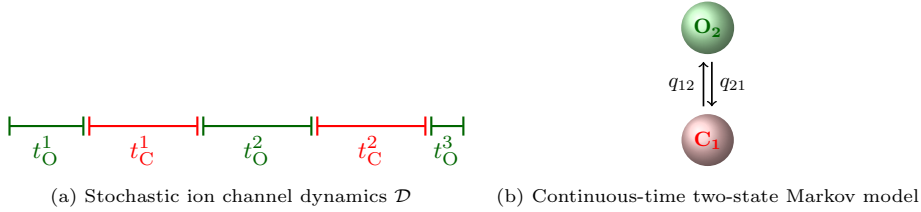

A simple model for the alternating sequence of open and closed
times~$\dyn$ is the two-state continuous-time Markov model which
describes stochastic transitions between a closed state~\rC[1] and an
open state~\gO[2]. A direct consequence of assuming that the
dynamics~$\dyn$ can be described by a Markov model is that the open
times~$\tO[i]$ and the closed times~$\tC[j]$ are
exponentially-distributed. Indeed, the Markov property requires that if the
model has been in a state~$S$ for a time~$t_0$, the probability that
it stays in this state for an additional duration~$\tau$ is
independent from the time~$t_0$ that has already passed. More
formally,

\begin{equation}
  \label{eq':markov}
  \prob(T > \tau + t_0) = \prob(T>t_0) \prob(T>\tau), \quad t_0 \geq
  0, \tau \geq 0 
\end{equation}

or, equivalently,

\begin{equation}
  \label{eq:nomemory}
  \prob(T > \tau + t_0 \, \vline \, T > t_0) = \prob(T>\tau), \quad t_0 \geq
  0, \tau \geq 0.
\end{equation}



The exponential distribution is the only continuous-time distribution
for which the ``lack of memory'' property~\eqref{eq:nomemory} holds. A
two-state Markov model can therefore just generate an alternating
sequence of exponentially-distributed sequence of open and closed
times. But exponentially-distributed open and closed time

distributions, $\fO(\tO)$ and $\fC(\tC)$, are insufficient for
describing the dynamics~$\dyn$ of many ion channels---a simple
counter-example is an ion channel that sometimes stays closed for a
short time and sometimes for a considerably longer
time. 
The most common approach for describing multiple timescales in the
open and closed time distributions is to introduce multiple open and
closed states. Intuitively, one might expect that each closed state
(or open state) represents a different exponentially distributed
timescale and each of these closed (or open) timescales is observed
with a specific probability. Indeed, as calculating the sojourn-time
distributions from a Markov model (with an arbitrary number of closed
and open states) shows, this expectation turns out to be true---open
and closed time distributions are mixture distributions of as many
exponential components as there are open/closed states, respectively
\citep{Col:81a}. Although the exponentials in the mixture
distributions~$\fO(\tO)$ and $\fC(\tC)$ can, in general, not be
identified with individual states in the Markov model, it is tempting
to interpret the set of open and closed states as a collection of
different timescales, ranging from ``slow'' to ``fast''.

Another attractive property of a Markov model consisting of multiple
open and closed states (a so-called \emph{aggregated Markov model}
with two classes) is that one might associate individual states of the
model with different biophysical states of the channel protein---one
well-known example is the de Young - Keizer model of the
inositol-trisphosphate receptor (\ipr) \citep{deY:92a}. To give a
simple example for this approach, consider a model that consists of a
chain of multiple closed states that eventually transition to an open
state, this is interpreted as transitions between various closed
conformations which eventually lead to an open state i.e. a
conformation in which the channel is open.

It is no coincidence that this idea has been highly influential---it
suggests that a model consisting of multiple open and closed states
can represent more than ``just'' the opening and closing of an ion
channel but, in addition, can reveal the hidden dynamics of
conformational changes underlying ``these'' dynamics~$\dyn$. For a
given number of open and closed states, $\nO$ and~$\nC$, there are
many different ways that these states can be connected to an
aggregated Markov model, see Section~\ref{sec:enumeration} for an
explicit calculation, the results are summarised in
Table~\ref{tab:enumeration}. Intuitively, it seems plausible that for
a given number of open and closed states, any of the different models
represent different dynamics~$\dyn$. A review of long-known results on
aggregated Markov models shows that this intuition turns out to be
clearly wrong, as we will explain in Section~\ref{sec:nonid}---models
with~$\nO$ open and~$\nC$ closed states can be reparametrised without
altering the dynamics~$\dyn$. For example, the fully-connected model
for any number of states~$\nO + \nC > 2$ can be continuously
reparametrised so that an infinite combination of parameters can
represent the same dynamics.

The phenomenon that a given model~$Q$ can be reparametrised to a
different model~$Q'$ that represents the dynamics~$\dyn$
equally well is known as non-identifiability. We will consider two
different aspects of this question:

\begin{enumerate}
\item For a given Markov model~$Q=(q_{ij})$ with $\nO$ open and $\nC$ closed
  states, can the rate constants~$q_{ij}$ 
  be uniquely identified from the dynamics~$\dyn$? If this is the
  case, the model~$Q$ is referred to as \emph{parameter-identifiable}.
\item 
  We now consider the \emph{model structure} i.e. the
  graph~$\mathcal{G}$ defined by the positive rates of an
  infinitesimal generator~$Q=(q_{ij})$. We ask the question if~$Q$ can
  be reparametrised to another model~$Q'$ so that they generate the
  same dynamics~$\dyn$ with a model structure~$\mathcal{G}'$ that is
  not equivalent to~$\mathcal{G}$ as a graph (by symmetries as
  described in Section~\ref{sec:enumeration}). We call two models with
  generators~$Q$ and~$Q'$ \emph{equivalent} if they produce the same
  dynamics~$\dyn$ despite being defined on different
  graphs~$\mathcal{G}$ and~$\mathcal{G}'$. If the dynamics~$\dyn$ can
  be generated by models defined on graphs~$\mathcal{G}$ and
  $\mathcal{G}'$ that are not equivalent we refer to this phenomenon
  as \emph{non-identifiability of model structure}.
\end{enumerate}


Since the seminal papers by \citet{Fre:85a, Fre:86a} it has been known
that models exceeding $2 \nO \nC$ rate constants can, in principle, be
reparametrised as in 1. i.e. these models lack parameter
identifiability. We provide an outline of the derivation of this
result by \citet{Fre:85a, Fre:86a} in Section~\ref{sec:fredkinrice} as
well as an alternative derivation in Section~\ref{sec:altproof}. As a
consequence, {we might restrict ourselves} to models
with fewer rate constants than this upper bound~$2 \nO \nC$. But even
models with rate constants below this bound can often be
re-parametrised to models that have the same number of rate constants
but whose states are connected in a different way (these models lack
identifiability of model structure as in 2.)---explicit examples are
presented in detail in Section~\ref{sec:nonidExample}. This means that
even in the best case that models cannot be continuously
reparametrised, there still is an equivalence class of a finite number
of models that represents the same dynamics~$\dyn$. This makes it
challenging to interpret aggregated Markov models as representations
of a particular biophysical mechanism---simply, because via
reparametrisation, the dynamics of a given model can be generated by a
different model that suggests a completely different mechanism.

We will then explore the implications of this non-identifiability of
model structure on models that are based on representations of
transitions between conformational states regulated by ligand binding
sites in Section~\ref{sec:biophysics}.

In the light of these difficulties, one might ask the question if
aggregated Markov models are a suitable structure for representing the
underlying biophysical mechanism of ion channels rather than just the
dynamics~$\dyn$. We have argued in~\citet{Sie:19b} that representing
biophysical mechanisms such as conformational changes is unlikely to
succeed by parametrising a Markov model from single channel data
obtained at steady-state ligand conditions alone. Instead, it is more
promising to obtain more direct evidence on the biophysical dynamics
by statistically analysing modal gating \citep{Ion:07a,Sie:14a} which
is known to be closely linked with conformational changes or using
additional data such as the response of the channel to rapid changes
of ligand concentrations \citep{Mak:07a}. In the Discussion
(Section~\ref{sec:discussion}) we will explain how hierarchical Markov
models \citet{Sie:16a} and Markov models with distributed delay
\citep{Haw:25a} can be used for representing aspects of the underlying
biophysical dynamics of the channel protein by providing structures
that can be parametrised directly with these additional data sources.


\section{Enumeration of ion channel models}
\label{sec:enumeration}

\begin{figure}[htbp]
  \centering
  \subfloat[]{
    \begin{minipage}{0.4\linewidth}
      \includegraphics[width=1.0\linewidth]{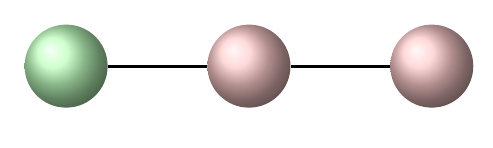}%
    \end{minipage}
    \begin{minipage}{0.1\linewidth}
      \centering
      vs.
    \end{minipage}
    \begin{minipage}{0.4\linewidth}
      \includegraphics[width=1.0\linewidth]{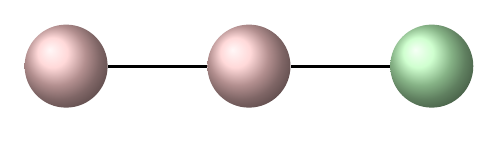}%
    \end{minipage}
  }%
      \\
    \subfloat[]{
      \begin{minipage}{0.4\linewidth}
        \centering
          \includegraphics[width=0.55\linewidth]{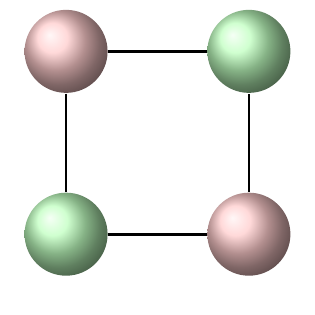}%
    \end{minipage}
    \begin{minipage}{0.1\linewidth}
      \centering
      vs.
    \end{minipage}
    \begin{minipage}{0.4\linewidth}
      \centering
      \includegraphics[width=0.55\linewidth]{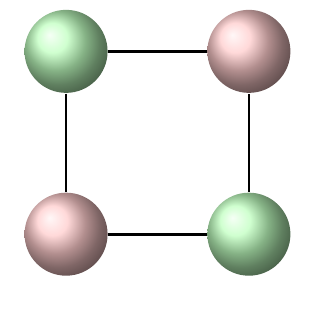}%
    \end{minipage}
  }%
  \\
    \subfloat[]{
    \begin{minipage}{0.4\linewidth}
      \includegraphics[width=0.9\linewidth]{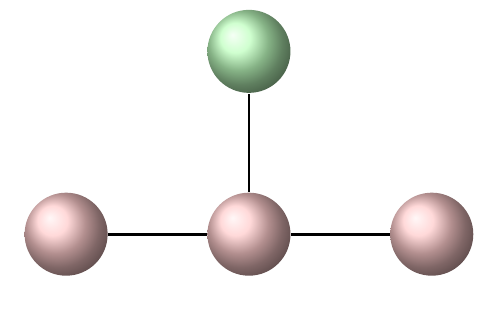}%
    \end{minipage}
    \begin{minipage}{0.1\linewidth}
      \centering
      vs.
    \end{minipage}
    \begin{minipage}{0.4\linewidth}
       \includegraphics[width=0.9\linewidth]{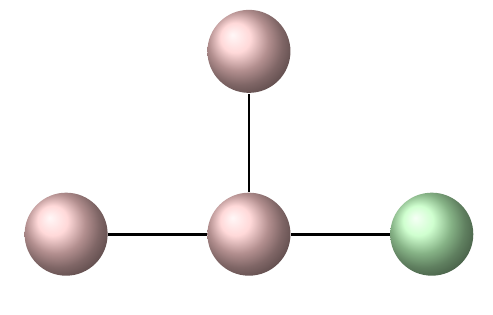}%
    \end{minipage}
  }%
  \caption{Three examples of graphs with vertices in two different
    colours which are equivalent (arguably, with increasing difficulty
    from (a) to (c)). ``Equivalent'' means that both are examples of
    the same unlabelled graph which has just been ``drawn'' in
    different ways. More formally, each graph on the left can be
    mapped to the graph on the right by swapping vertices without
    changing the adjacency structure. For a given graph which vertices
    can be mapped to each other is determined by the symmetry group of
    the graph. Enumeration of graphs with given properties relies on
    being able to find the symmetry group of this class of graphs and
    calculating the so-called cycle index, a polynomial which allows
    to keep track of the different equivalence classes under the
    operation of the symmetry group.}
  \label{fig:graphs}
\end{figure}

In \citet{Bru:05a}, the number of graphs with $\nO$ open and $\nC$
closed states has been stated (and credited to David Torney) but
further information about the derivation of these numbers have not
been given. Here, the calculation of the number of graphs is
explained. To illustrate the challenge we show a few examples of
graphs that are equivalent, although this is not obvious at first
glance, see Figure~\ref{fig:graphs}.

\subsection{P\'{o}lya enumeration}
\label{sec:polya}

We will ``count'' the number of different graphs using a technique
called \emph{P\'olya enumeration}\footnote{Although not the original
  developer of the theory, P\'{o}lya has been very influential in its
  popularisation and extension in the seminal article
  \citep{Pol:37a}.}. Behind P\'{o}lya enumeration is the idea that the
number of objects with a particular property can be represented by
polynomials $p(t)$.

Let us first look at the very simple example of $k$-subsets of a set
of $n$ elements. If $1$ stands for the empty set~$\emptyset$ and $t$
for the set~$\mathcal{S}_1 = \{1\}$ with one element, the polynomial

\begin{equation}
  \label{eq:1set}
  1 + t
\end{equation}

represents the two possible subsets with zero and one element. One
interpretation of the polynomial~$1+t$ is that we can
either draw the (single) element $1$ or not.

For the set~$\mathcal{S}_2=\{ 1, 2 \}$ we can extend this idea by
considering the product

\begin{equation}
  \label{eq:2set}
\underbrace{(1+t)}_{\text{draw $1$?}} \cdot
\underbrace{(1+t)}_{\text{draw $2$?}} 
\end{equation}

where the first factor~$(1+t)$ stands for drawing the element $1$ (or not) and
the second factor stands for drawing element~$2$ (or not). It is clear
that the polynomial

\[
 1 \cdot 1 + 1 \cdot t + t \cdot 1 + t \cdot t = 1 + 2 t + t^2
\]

enumerates all possible outcomes of drawing (without replacement)
from~$\mathcal{S}_2$. After simplifying the polynomial, the term $1$
stands again for the empty set~$\emptyset$, $2 t$ represents the
subsets $\{1\}$ and~$\{2\}$ with one element each and $t^2$ accounts
for the full set~$\mathcal{S}_2=\{1, 2 \}$ with two elements. Going
forward with this idea, for the set~$\mathcal{S}_n=\{ 1, \dots, n\}$
with $n$ elements, the process of drawing (or not) each of the $n$
elements of the set can be represented via the polynomial

\begin{equation}
  \label{eq:nset}
(1 +t)^n = \underbrace{(1+t)}_{\text{draw $1$?}} \cdot
\underbrace{(1+t)}_{\text{draw $2$?}} \cdots \underbrace{(1+t)}_{\text{draw $n$?}}
\end{equation}

and via binomial expansion of~\eqref{eq:nset} we obtain 

\begin{equation}
  \label{eq:1}
  (1 +t)^n = \sum_{k=0}^n
  \begin{pmatrix}
    n\\k
  \end{pmatrix} t^k
\end{equation}

i.e. indeed the coefficient~$
\begin{pmatrix}
  n \\ k
\end{pmatrix}$ of each term~$t^k$ indicates the number of $k$-subsets
in a set with~$n$ elements. 

In summary, the laws of polynomial arithmetic ensure that all
possibilities of drawing~$k$ elements from a set with~$n$ elements are
summarised appropriately.

\subsection{Enumerating graphs}
\label{sec:graphs}

\begin{figure}[htbp]
  \subfloat[]{%
    \includegraphics[width=0.15\linewidth]{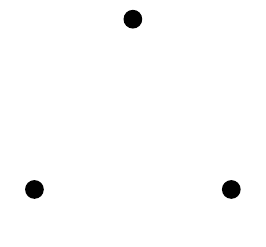}%
  }\\
   \subfloat[]{%
        \includegraphics[width=0.15\linewidth]{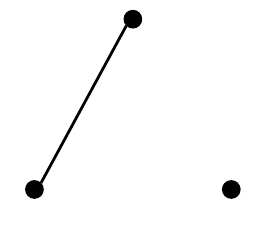}%
      }%
      \qquad
       \subfloat[]{%
        \includegraphics[width=0.15\linewidth]{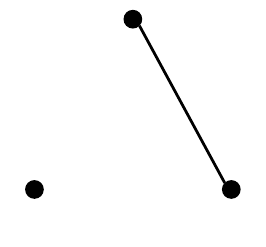}%
      }%
      \qquad
       \subfloat[]{%
        \includegraphics[width=0.15\linewidth]{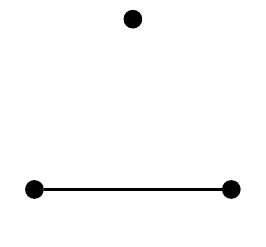}%
      }\\
      \subfloat[]{%
        \includegraphics[width=0.15\linewidth]{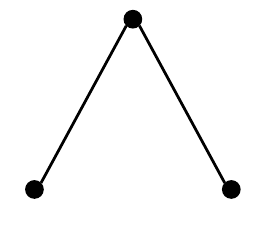}%
      }%
      \qquad
      \subfloat[]{%
        \includegraphics[width=0.15\linewidth]{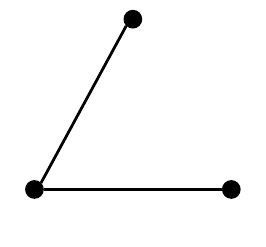}%
      }%
      \qquad
      \subfloat[]{%
        \includegraphics[width=0.15\linewidth]{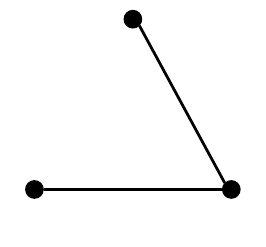}%
      }\\
      \subfloat[]{%
        \includegraphics[width=0.15\linewidth]{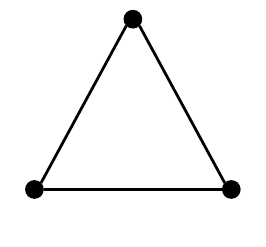}%
      }
  \caption{Graphs with three vertices. It is easy to see that, unless
    the graph is labelled, in each row, all graphs are equivalent to
    the first one.}
  \label{fig:G3}
\end{figure}

We now apply the idea explained in the previous section to a method
for calculating the number of unlabelled graphs for a given number of
vertices\footnote{The application of P\'{o}lya enumeration to graphs
  is presented in the monograph by \citet{Har:73a}.}. A graph~$\mathcal{G}$
with~$\nV$ vertices has a maximal number of~$\nE=
\begin{pmatrix}
  \nV \\
  2
\end{pmatrix}$ edges. Similar to the calculation of the $k$-subsets of
the set $\mathcal{S}_n$ with~$n$ elements, the idea of enumerating
graphs is based on drawing without replacement from the edge
set. Unfortunately, the sequence of coefficients~$
\begin{pmatrix}
  \nE \\k
\end{pmatrix}$ of the polynomial~$(1+t)^{\nE}$ does not yield correct
results for graphs with more than two vertices. For example, for a
graph with~$\nV=3$ vertices (which has up to~$\nE=3$ edges), the
formula yields~$1 + 3 t + 3 t^2 + t^3$, indicating that there are 3
graphs with one edge and 3 graphs with two edges. This is the correct
result for a labelled graph i.e. a graph where any pair of vertices
(and thus, also edges between them) can be distinguished from each
other. If vertices are not labelled, any rearrangement of vertices
that leaves the edge structure of the graph unchanged, is considered
to represent the same graph, see Fig.~\ref{fig:G3}.

\subsubsection{Symmetry group of a graph---the line group}
\label{sec:symmetry}

\begin{figure}[htbp]
  \subfloat[(1)(2)(3)]{%
    \label{fig:(1)(2)(3)}
    \includegraphics[width=0.3\linewidth]{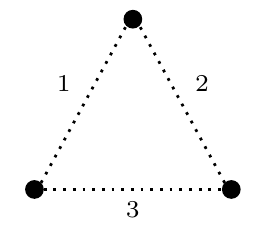}%
  }%
  \\
  \subfloat[(1)(23)]{%
    \label{fig:(1)(23)}
    \includegraphics[width=0.3\linewidth]{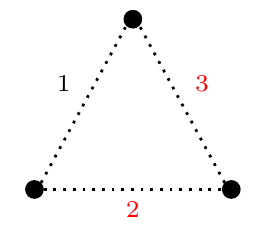}%
  }%
  \;
  \subfloat[(2)(13)]{%
    \label{fig:(2)(13)}
    \includegraphics[width=0.3\linewidth]{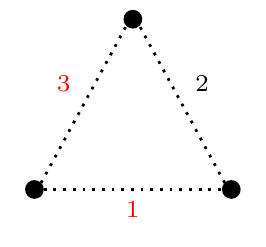}%
  }%
  \;
  \subfloat[(3)(12)]{%
    \label{fig:(3)(12)}
    \includegraphics[width=0.3\linewidth]{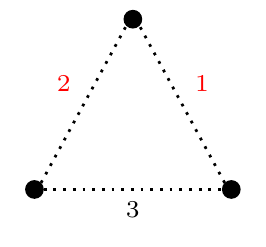}%
  }%
  \\
  \subfloat[(123)]{%
    \label{fig:(123)}
    \includegraphics[width=0.3\linewidth]{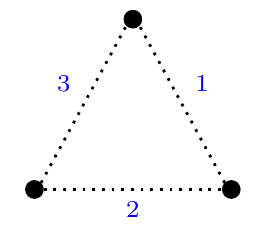}%
  }%
  \;
  \subfloat[(132)]{%
    \label{fig:(132)}
    \includegraphics[width=0.3\linewidth]{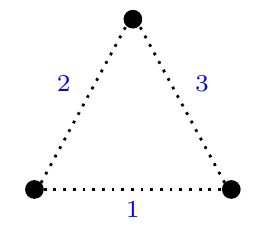}%
  }%
  \caption{Representation of the six different permutations of the
    graph with three vertices.}
  \label{fig:S3}
\end{figure}

Because we enumerate graphs by the number~$\nE$ of edges we consider
the symmetry group on the edge set of the graph, the line group. As an
example, the symmetries of a graph with~$\nV=3$ vertices are shown in
Fig.~\ref{fig:S3}.
Each symmetry is a permutation of the edges of the
graph and we will represent each permutation via disjoint
cycles. Within a cycle~$(\cdots)$, each element is mapped to the
subsequent element and the last element is mapped to the first one
that appears in the cycle (hence the name ``cycle''). For example, the
cycle~$(1 3 2)$ maps edge $1$ to edge $3$, edge $3$ to edge $2$ and
edge $2$ to edge $1$, see Fig.~\ref{fig:(132)}. The cycle $(2 3)$
swaps edges $2$ and $3$, in order to account for what happens to edge
$1$, we need to add the cycle~$(1)$ to account for the fact that edge
$1$ is left unchanged so that we obtain the cycle
representation~$(1)(23)$ for the permutation shown in
Fig.~\ref{fig:(1)(23)}. It is well-known that any permutation can be
represented via \emph{disjoint cycles} i.e. each element appears in
any of the cycles exactly once. For a given cycle~$(\cdots)$, the
number of elements that appear within the cycle are the length of the
cycle. For example, the permutation~$(1) (23)$ consists of one cycle
of length~$1$ and one cycle of length~$2$. 

\subsubsection{The cycle index}
\label{sec:cycleindex}

The disjoint cycle representation is important because it will help us
construct a multivariate polynomial, known as the \emph{cycle index}
which will enable us to account for the effect of permutations on the
number of elements in a set after identifying equivalent elements. The
cycle index is usually introduced in a very abstract form which makes
it applicable to a wide range of applications but makes it difficult
for those unfamiliar with the general theory to understand how it is
applied in a particular situation. For our purpose of understanding
the specific example of enumerating graphs we hope that the reader
will find it simpler to consider the concrete example of a graph with
three vertices to understand the principle of the construction of the
cycle index. A general presentation of the theory can be found in
\citet{Pol:37a,Har:73a}.

The cycle index represents each permutation that has been written in
the disjoint cycle representation via a multivariate monomial
$t_1^{\alpha_1} t_2^{\alpha_2} \dots\ t_n^{\alpha_n}$. Here, each
variable $t_i$ stands for a cycle of length~$i$ and the
exponent~$\alpha_i$ accounts for how many times a cycle of length~$i$
occurs in a given permutation. This means that, for example, the
permutation~$(1)(2)(3)$ i.e. the identity is represented as~$t_1^3$,
for $(1)(23)$ we obtain $t_1 t_2$ and for $(132)$ we have~$t_3^1$.

The benefit of introducing the variables~$t_i$ is that they allow us
to account for the effect of a cycle of length~$i$. Consider, for
example, the permutation~$(1)(23)$, see
Fig.~\ref{fig:(1)(23)}. Because edges 2 or 3 are swapped, the
resulting graph can only be equivalent if both edges are either
present or absent i.e. for a cycle of length 2 we need to either draw
none of the edges in the circle or both. This can be represented by
the polynomial $1+t^2$ where~$t^2$ represents drawing two
edges. Because edge~$1$ is not moved we can either include this edge
in the graph or not which is represented, as above, as the
polynomial~$1+t$. In general, to see the effect of a cycle of
length~$i$ we replace the variable~$t_i$ with $1+t^i$.

To account for multiple permutations with a particular combination of
cycles---these permutations are said to have the same type---we
multiply each monomial with the number of permutations of the type
represented by this monomial. 

The cycle index of the line group~$\mathfrak{S}^{(2)}_3$ of graphs
with~$\nV=3$ vertices is

\begin{equation}
  \label{eq:cycInd3}
  Z(\mathfrak{S}^{(2)}_3; t_1, t_2, t_3) = \frac{1}{6} \left( t_1^3 + 3 t_1 t_2 + 2 t_3 \right)
\end{equation}

where~$t_1^3$ accounts for the identity, $3 t_1 t_2$ stands for the 3
permutations that swap two edges while not moving the remaining one
and $2 t_3$ represents the two permutations that move all three edges,
see Fig.~\ref{fig:S3}. The polynomial is scaled by the number of
permutations in the symmetry group which results in the
factor~$\frac{1}{6}$.

According to P\'{o}lya's enumeration theorem, by evaluating the cycle
index~$Z(\mathfrak{S}^{(2)}_3, 1+t, 1+t^2, 1+t^3)$ we can find a
polynomial that represents the number of graphs with 3 vertices for
different numbers of edges:

\begin{align}
  \label{eq:9}
 Z(1+t, 1+t^2, 1+t^3) &= \frac{1}{6} \left[  (1+t)^3 + 3\cdot (1+t)(1+t^2)
                         + 2\cdot(1+t^3) \right] \\
                   &= 1 + t + t^2 +t ^3.
\end{align}

This result indicates that among graphs with three vertices we can
find one unlabelled graph each with zero, 1, 2 or 3 edges. This is
consistent with Fig.~\ref{fig:G3}.

Based on this example, we can conclude that P\'{o}lya enumeration
relies on finding an expression of the cycle index for the symmetry
group for a particular class of objects of interest. This is not an
easy task and leads to a quite complicated closed-form formula in the
general case. Because the main aim of this section is to explain the
underlying idea of the enumeration of aggregated Markov models rather
than explaining the calculation of the line group~$\mathfrak{S}^{(2)}_n$in full detail we refer the reader to
chapter 4.1 in \citet{Har:73a}. The number of graphs for a given
number of vertices~$\nV$, calculated in this way can be found in
Table~\ref{tab:enumeration}.

\subsection{Graphs with coloured vertices and rooted graphs}
\label{sec:rootedgraphs}

\begin{figure}[htbp]
  \centering
  \subfloat[Rooted graphs]{%
    \begin{minipage}{0.15\linewidth}
      \includegraphics[width=1.0\linewidth]{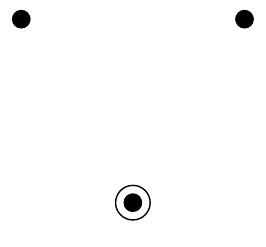}%
    \end{minipage}
   ~\vline ~
    \begin{minipage}{0.15\linewidth}
      \includegraphics[width=1.0\linewidth]{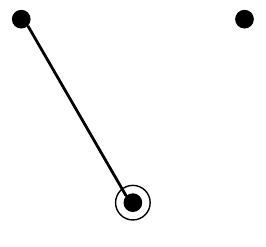}%
    \end{minipage}
         ~\vline~
          \begin{minipage}{0.15\linewidth}
            \includegraphics[width=1.0\linewidth]{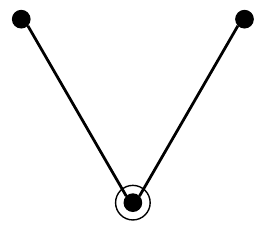}%
          \end{minipage}
          ~\vline ~
          \begin{minipage}{0.15\linewidth}
            \includegraphics[width=1.0\linewidth]{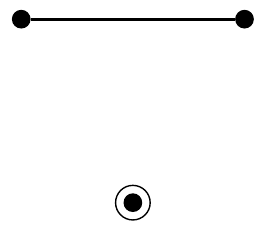}%
          \end{minipage}
          ~\vline ~
          \begin{minipage}{0.15\linewidth}
            \includegraphics[width=1.0\linewidth]{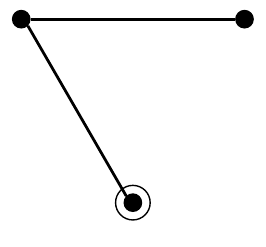}%
          \end{minipage}
          ~\vline ~
          \begin{minipage}{0.15\linewidth}
            \includegraphics[width=1.0\linewidth]{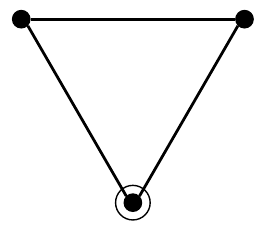}%
          \end{minipage}
        }%
        \\
        \subfloat[Coloured graphs]{%
          \begin{minipage}{0.15\linewidth}          
            \includegraphics[width=1.0\linewidth]{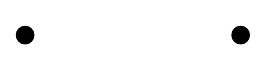}%
          \end{minipage}
          ~\vline ~
          \begin{minipage}{0.15\linewidth}
            \includegraphics[width=1.0\linewidth]{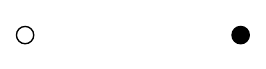}%
          \end{minipage}
          ~\vline ~
          \begin{minipage}{0.15\linewidth}
            \includegraphics[width=1.0\linewidth]{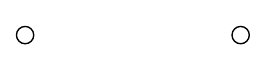}%
          \end{minipage}
          ~\vline ~
          \begin{minipage}{0.15\linewidth}
            \includegraphics[width=1.0\linewidth]{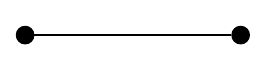}%
          \end{minipage}
          ~\vline ~
          \begin{minipage}{0.15\linewidth}
            \includegraphics[width=1.0\linewidth]{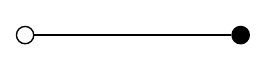}%
          \end{minipage}
          ~\vline ~
          \begin{minipage}{0.15\linewidth}
            \includegraphics[width=1.0\linewidth]{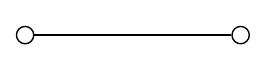}%
          \end{minipage}
        }%
        \caption{Rooted graphs with 3 vertices and the corresponding
          coloured graphs with 2 vertices. Vertices that are adjacent
          to the root are represented as white in the corresponding
          coloured graph, vertices that are not adjacent are black.}
  \label{fig:rootG}
\end{figure}

Enumerating aggregated Markov models comes with the additional
difficulty that the vertices of the graph can have two different
``colours'', say red and green, as in Fig.~\ref{fig:G3}. An elegant
solution for enumerating graphs with two colours is to consider rooted
graphs, see Figure~\ref{fig:rootG}.

``Colour'' is represented in a rooted graph by considering if a
designated root node is adjacent to a vertex or not---in
Fig.~\ref{fig:rootG}, a vertex which is adjacent to the root is
coloured white, if it is not adjacent to the root it is coloured
black. Representing the colour of a vertex as part of the structure of
the graph is much simpler than attempting to investigate the different
colourings of graphs under their respective symmetry groups. The
calculation of the cycle index for rooted graphs is again described in
\citet[chapter 4.4]{Har:73a}, results for the number of rooted graphs
for a given number of vertices~$\nV$ are found in
Table~\ref{tab:enumeration}.

\subsection{Connected rooted graphs\dots\ and The End}
\label{sec:connected}

In our attempt to enumerate the number of aggregated Markov models
there is one difficulty that remains. Instead of all graphs whose
vertices can have two colours we are, in fact, only interested in
connected graphs. There is a general method for determining the number
of connected graphs from the number of graphs with a given
property. For this, we consider the formal power series (also known as
the generating series)

\begin{equation}
  \label{eq:g}
  g(t) = \sum_{k=1}^{\infty} g_k t^k
\end{equation}

where $g_p$ is the number of graphs with a certain property, for
example, the number of arbitrary graphs or the number of rooted graphs
with~$k$ vertices---note that in contrast to the examples for
enumeration of graphs and rooted graphs considered above we are not
enumerating graphs by the number of edges.

If we define, analogously to~\eqref{eq:g}, the generating
series~$c(t)$ of connected graphs or rooted graphs, $g(t)$ and $c(t)$
are related by \emph{Riddell's formula}\footnote{In terms of integer
  sequences the relationship~\eqref{eq:riddell} is also known as an
  Euler transform.} \citep{Rid:51a, Har:73a}:

\begin{equation}
  \label{eq:riddell}
  1 + g(t) = \exp \left( \sum_{k=1}^\infty\frac{1}{k} c(t^k) \right).
\end{equation}

This relationship between graphs
and connected graphs might look rather mysterious but it is based on
the simple fact that for any graph we can determine the number of
connected components. Thus, the number of graphs can be reconstructed
by drawing connected graphs that will then become the connected
components of a regular graph. The additional term~$+1$
in~\eqref{eq:riddell} stands for the ``empty'' graph with no
vertices. The relevant symmetry group for graphs consisting of~$n$
connected components is~$\mathfrak{S}_n$ so that we obtain the cycle index

\begin{equation}
  \label{eq:cycIndConn}
  \sum_{k=1}^{\infty} Z(\mathfrak{S}_n; c(t))  =: Z(S_\infty, c(t)).
\end{equation}

It can be shown that as a formal power series,

\begin{equation}
  \label{eq:expConn}
  Z(\mathfrak{S}_\infty; c(t)) = \exp \left( \sum_{k=1}^\infty\frac{1}{k} c(t^k) \right).
\end{equation}

By taking advantage of the arithmetic of formal power
series---although this involves some tedious calculations---it is
possible to use~\eqref{eq:riddell} for calculating the generating
series~$c(t)$ for connected graphs from the generating series~$g(t)$
for arbitrary graphs. This process is also known as an inverse Euler
transform of~\eqref{eq:riddell}. 

One possible solution is to set

\begin{equation}
  \label{eq:logPowerSeries}
  \sum_{k=1}^\infty a_k t^k = \log ( 1 + g(t))
\end{equation}

and calculate the coefficients~$a_k$ by coefficient matching
with~$\log ( 1 + g(t))$ as a power series. This yields the recursion
formula

\begin{equation}
  \label{eq:recA}
  a_k = g_k - \frac{1}{k} \sum_{i=1}^{k-1} i \cdot  a_i g_{k-i}.
\end{equation}

Matching coefficients of~$\sum_{k=1}^\infty a_k t^k$
with~$\sum_{k=1}^\infty\frac{1}{k} \cdot c(t^k)$ leads to the
relationship

\begin{equation}
  \label{eq:cKfromaK}
  a_k = \frac{1}{k} \sum_{d | k} d c_d.
\end{equation}

Finally, the coefficients~$c_k$ of the generating series for connected
graphs can be calculated from~$a_k$ by \emph{M\"obius inversion}:

\begin{equation}
  \label{eq:cK}
  c_k = \sum_{d | k} \frac{\mu(d)}{d} a_{p/d}
\end{equation}

where $d | k$ stands for all divisors~$d$ of~$k$ and the M\"obius
function is defined as

\begin{equation}
  \label{eq:moebius}
  \mu(k) = \left\{
      \begin{array}{ll}
        1 & \text{ if } n=1,\\
        (-1)^k & \text{ if } n \text{ is a product of } k \text{
                 distinct primes},\\
        0 & \text{ if } n \text{ is divisible by a square}.
      \end{array}
    \right.
\end{equation}

Using the formulae~\eqref{eq:recA} and~\eqref{eq:cK} it is
straightforward to calculate the number of connected graphs as well as
the number of rooted graphs, see Table~\ref{tab:enumeration}.

The number of connected rooted graphs with~$\nV+1$ vertices (including
the root node) or, analogously, graphs with~$\nV$ vertices which have
one of two different colours, also contains graphs that have only one
of the two colours. Those graphs are examples for degenerate aggregated
Markov models where only one of the two aggregates is
present. Fortunately, it is not difficult to exclude these
graphs---for every connected graph there are only two graphs
that have only one colour---one in each colour. Thus, the number of (non-degenerate)
aggregated Markov models~$M_k$ with~$\nV$ vertices is obtained from the
number of coloured graphs~$CR_k$ by subtracting twice the number of
connected graphs~$2 c_k$:

\begin{equation}
  \label{eq:AMM}
  M_k = CR_k - 2\cdot c_k.
\end{equation}

\begin{table}
  \centering
  \begin{tabular}{*{6}r}
    $\nV$ & Graphs & Connected graphs & Rooted graphs & Connected rooted graphs &
                                                             Aggregated
                                                             Markov models\\
    \midrule
    1 & 1 & 1 & 2 & 2 & 0\\
    2 & 2 & 1 & 6 & 3 & 1\\
    3 & 4 & 2 & 20 & 10 & 6\\
    4 & 11 & 6 & 90 & 50 & 38\\
    5& 34 &  21 & 544 & 354 & 312 \\
    6 & 156 & 112 & 5096 & 3883 & 3659\\
    7& 1044& 853 & 79264&   67994 &66288 \\
    8 & 12346& 11117 & 2208612& 2038236 & 2016002 \\
    9 &274668 & 261080 & 113743760& 109141344 & 108619184 \\
    10 &12005168 &11716571 & 10926227136& 10693855251 & 10670422109
  \end{tabular}
  \caption{Enumeration of graphs, rooted graphs, connected rooted
    graphs and aggregated Markov models obtained by P\'{o}lya
    enumeration as explained in Section~\ref{sec:enumeration}. Note
    that for rooted and connected rooted graphs we have not counted
    the root node i.e. any rooted graphs with~$\nV$ vertices
    has~$\nV+1$ vertices including the root.}
  \label{tab:enumeration}
\end{table}




\section{Non-identifiability}
\label{sec:nonid}

One aspect of aggregated Markov models is that non-identifiability is
a common phenomenon. Non-identifiability can be illustrated by the
hypothetical situation that we know exactly the structure of the
underlying model that generates the process that we observe. If the
model is non-identifiable, it can generate identical behaviour when
choosing different parameter
sets. 
The reason why such a model is called non-identifiable can be
explained by the somewhat counter-intuitive consequence that even if
an infinite amount of data which is not perturbed by noise were
available, the parameters of the model nevertheless could not be
uniquely determined by fitting the model to these data.

\subsection{Upper bound for the number of parameters in aggregated
  Markov models}
\label{sec:fredkinrice}

The main result in this regard goes back to \citet{Fre:85a,Fre:86a}
who observed that the dynamics of aggregated Markov models can be
completely represented by the bivariate distributions $\fOC$ and
$\fCO$ of the length of an open time $\tO$ followed by a closed time $\tC$ and
vice versa :
\begin{align}
  \label{eq:fOC}
       \fOC (\tO, \tC) &= \sum_{i=1}^{\nO}\sum_{j=1}^{\nC}
        \alpha_{\text{\gO \rC}}^{ij} \exp\left(-\lamO[i] \tO -
                         \lamC[j]  \tC
                         \right) \\
  \label{eq:fOCrho}
                       &= \sum_{i=1}^{\nO}\sum_{j=1}^{\nC}
        \rho_{\text{\gO \rC}}^{ij} \lamO[i] \lamC[j] \exp\left(-\lamO[i] \tO -
                         \lamC[j]  \tC
                                       \right),\\
  \label{eq:fCO}
                                     f_{\text{\rC \gO}}
  (t_{\text{\rC}},  t_{\text{\gO}}) &= \sum_{j=1}^{\nC}\sum_{i=1}^{\nO}
        \alpha_{\text{\rC\gO }}^{ji} \exp\left(-\lamC[j] \tC -
                                      \lamO[i]  \tO \right),\\
                       &= \sum_{j=1}^{\nC}\sum_{i=1}^{\nO}
        \rho_{\text{\rC\gO }}^{ji} \lamC[j] \lamO[i]  \exp\left(-\lamC[j] \tC -  \lamO[i]  \tO \right).
      \end{align}

      Note that here, the parameters~$\lamO[i]$
      and~$\lamC[j]$ are the negatives of eigenvalues of submatrices of
      the infinitesimal generator~$Q$ of the aggregated Markov
      model. 

      The bivariate distributions~\eqref{eq:fOC}, \eqref{eq:fCO} can
      also be represented as bilinear forms. By defining the matrices

      \begin{align}
        \label{eq:Aoc}
        \boldsymbol{\alpha}_{\text{\gO \rC}} &= \left( \alpha_{\text{\gO
                                      \rC}}^{ij} \right) \\
                \label{eq:Aco}
        \boldsymbol{\alpha}_{\text{\rC \gO}} &= \left( \alpha_{\text{\rC \gO}}^{ji} \right)
      \end{align}

      we obtain

      \begin{align}
                \label{eq:fOCvect}
        \fOC (\tO, \tC) &= \left(\exp(-\lamO[1] \tO), \exp(-\lamO[2] \tO),
                          \dots, \exp(-\lamO[\nO] \tO) \right)
                         \boldsymbol{\alpha}_{\text{\gO \rC}}
                          \begin{pmatrix}
                            \exp(-\lamC[1] \tC)\\
                            \exp(-\lamC[2] \tC)\\
                            \vdots \\
                            \exp(-\lamC[\nC] \tC)
                          \end{pmatrix} \\
        \label{eq:fCOvect}
        \fCO (\tC, \tO) &=\left(\exp(-\lamC[1] \tC), \exp(-\lamC[2] \tC),
                          \dots, \exp(-\lamC[\nC] \tC) \right)
                         \boldsymbol{\alpha}_{\text{\rC \gO }}
                          \begin{pmatrix}
                            \exp(-\lamO[1] \tO)\\
                            \exp(-\lamO[2] \tO)\\
                            \vdots \\
                            \exp(-\lamO[\nO] \tO)
                          \end{pmatrix}
      \end{align}

      Using the result from \citep{Fre:85a,Fre:86a} that all
      information on the dynamics of an aggregated Markov model is
      contained in the bivariate distributions~$\fOC$ and~$\fCO$
      (Theorem 4.1 in \citet{Fre:85a}, and Theorem A in
      \citet{Fre:86a})\footnote{More formally, the statement that ``all
        information on the dynamics of an aggregated Markov model is
        contained in the bivariate distributions~$\fOC$ and~$\fCO$''
        means that the parameters of higher-order distributions such
        as $f_{\text{\gO\rC\gO}}$ or $f_{\text{\rC\gO\rC}}$ are
        uniquely determined by the bivariate distributions~$\fOC$
        and~$\fCO$. See \citet{Fre:85a,Fre:86a}.}, we can find the maximal number of free
      parameters of an aggregated Markov by answering the simpler
      question how many parameters the distributions~$\fOC$ and~$\fCO$
      depend on. The numbers of free parameters in the univariate
      distributions $\fO$ and $\fC$ are contained in this number
      because these distributions can be obtained by marginalising the
      bivariate distributions, for example,

\begin{align}
  \label{eq:fOfromfOC}
  \fO(\tO)  = \int_0^\infty \fOC (\tO, \tC) d\tC & =\sum_{j=1}^{\nC}\sum_{i=1}^{\nO}
  \frac{\alpha_{\text{\gO \rC}}^{ij}}{\lamC[j]} \exp\left(-\lamO[i] \tO
  \right),\\
  &=\sum_{i=1}^{\nO} \underbrace{\left( \sum_{j=1}^{\nC}
  \frac{\alpha_{\text{\gO \rC}}^{ij}}{\lamC[j]} \right)
    }_{\alpha^i_{\text {\gO}}}\exp\left(-\lamO[i] \tO
  \right),
\end{align}


To determine the number of parameters contained in the
distributions~$\fOC$ and $\fCO$\footnote{Note that by considering the
  distributions~$\fOC$ and~$\fCO$, the upper bound for parameters
  derived by \citet{Fre:85a,Fre:86a} is only valid for processes that
  are at equilibrium. Recordings from single ion channels are usually
  started after the channels had sufficient time to adjust to the
  experimental conditions such as, for example, ligand concentrations. For
  this reason, the assumption that the ion channel dynamics~$\dyn$ has
  reached equilibrium is justified. A theory for the
  non-identifiability of aggregated Markov models that have not
  reached the stationary distribution has been developed by
  \citet{Ito:92a} for discrete-time aggregated Markov models and was
  then transferred to the continuous case by \citet{Ryd:96a}.} we
follow the derivation given in \citet[Corollary 4.1]{Fre:85a}.

To find the total number of free parameters contained in~$\fOC$ and
$\fCO$ we consider~\eqref{eq:fCO} or~\eqref{eq:fOCvect} and observe
that we have to account for the $\nO$ rates~$\lamO[i]$ and the $\nC$
rates~$\lamC[j]$ and add them to the number of parameters contained
in~$\alpha^{ij}_{\text{\gO\rC}}$ and~$\alpha^{ji}_{\text{\rC
    \gO}}$. Considering that the $\alpha^{ij}_{\text{\gO\rC}}$
and~$\alpha^{ji}_{\text{\rC \gO}}$ can be interpreted as tables
with~$\nO \cdot \nC$ parameters each, the number of parameters that
both~$\alpha^{ij}_{\text{\gO\rC}}$ and~$\alpha^{ji}_{\text{\rC \gO}}$
depend on appears to be~$2\nO \nC$ so that the total number of
parameters would be $2\nO \nC + \nO + \nC$. But there are a few
constraints that reduce this number.

First of all, considering only one of the distributions, say~$\fOC$,
we observe that the number of parameters is reduced by
one---because~$\fOC$ is a probability distribution, the
~{$\rho^{ij}_{\text{\gO\rC}}$, see~\eqref{eq:fOCrho},} have to sum up to one. This
yields~$\nO \nC -1$ free parameters.

Second, 
there are constraints that arise from the coupling
of~$\alpha^{ij}_{\text{\gO\rC}}$ and~$\alpha^{ji}_{\text{\rC \gO}}$
introduced by the fact that~$\fO$ and $\fC$ can both be obtained from
the bivariate distributions~$\fOC$ and~$\fCO$ via
marginalisation. Calculating~$\fO$ by marginalising~$\fOC$ as shown in
\eqref{eq:fOfromfOC} is just one of two possibilities---$\fO$ can also
be calculated analogously to~\eqref{eq:fOfromfOC} from the
distribution~$\fCO$. Whilst marginalising $\fOC$ yields coefficients
$\alpha^i_{\text{\gO}} = \sum_{j=1}^{\nC} \frac{\alpha_{\text{\gO \rC}}^{ij}}{
  \lamC[j]}$, marginalising~$\fCO$ leads to
coefficients~$\alpha^i_{\text{\gO}}$ that depend on
the~$\alpha_{\text{\rC\gO}}^{ji}$ instead. Because calculating the
coefficients~$\alpha^i_{\text {\gO}}$ from~$\fCO$ instead of~$\fOC$
must not change their value, we obtain the constraints

\begin{equation}
  \label{eq:aCOCons}
\sum_{j=1}^{\nC} \frac{\alpha_{\text{\rC \gO}}^{ji}}{
  \lamC[j]} =  \sum_{j=1}^{\nC} \frac{\alpha_{\text{\gO \rC}}^{ij}}{
  \lamC[j]}, \quad i=1, \dots, \nO
\end{equation}

and analogously, by considering the calculation of~$\fC$ via
marginalisation of $\fOC$ and $\fCO$, respectively, further constraints
\begin{equation}
  \label{eq:aOCCons}
\sum_{i=1}^{\nO} \frac{\alpha_{\text{\rC \gO}}^{ji}}{
  \lamO[i]} =  \sum_{i=1}^{\nO} \frac{\alpha_{\text{\gO \rC}}^{ij}}{
  \lamO[i]}, \quad j=1, \dots, \nC
\end{equation}

on the~$\alpha^{ji}_{\text{\rC\gO}}$ are
obtained. Interpreting~$\alpha^{ji}_{\text{\rC\gO}}$ as a
$\nC \times \nO$ matrix, these constraints allow us to calculate one
element in each column from the equations in~\eqref{eq:aCOCons} and,
similarly, using~\eqref{eq:aOCCons}, one element in each row
of~$\left(\alpha^{ji}_{\text{\rC\gO}} \right)$ from the
$\alpha_{\text{\gO\rC}}^{ij}$ and other entries in the same row or
column. This means that the $\nC \times \nO$ matrix
~$\left(\alpha^{ji}_{\text{\rC\gO}} \right)$ effectively has
only~$(\nO - 1)(\nC - 1)$ components. Thus, in total counting the
number of independent parameters contained in the coefficients
$\alpha^{ij}_{\text{\gO\rC}}$ and~$\alpha^{ji}_{\text{\rC \gO}}$ we
conclude that there are

\begin{equation}
  \label{eq:totalAoc}
  \nO \nC - 1 + (\nO-1)(\nC-1) = 2 \nO \nC - \nO - \nC
\end{equation}

Now, taking into account that there are~$\nO$ rate
constants~$\lamO[i]$ and $\nC$ rate constants~$\lamC[j]$ we find that
in total there are at most~$2 \nO \nC$ independent parameters in the
distributions~$\fOC$ and~$\fCO$.

We note that a refined upper bound for the maximum number of rates of
an aggregated Markov model has been derived in \citet{Fre:86a}. Apart
from the numbers of open and closed, \citet{Fre:86a} also considered
the rank of the transmission matrices~$Q_{\text{\rC \gO}}$
and~$Q_{\text{\rC \gO}}$. For more details we refer the reader to
Lemma A in \citet{Fre:86a} where this result is presented.





\subsection{Alternative derivation of the upper bound~$2 \nO \nC$}
\label{sec:altproof}

We would like to derive the upper bound $2\nO \nC$ for the number of
rate constants of an aggregated Markov model more directly, as
presented previously in similar form in
\citep{Kie:89a}. For this purpose it is useful to write the
infinitesimal generator~$Q$ of an aggregated Markov in the form

\begin{equation}
  \label{eq:Q}
  \mathbf{Q} =
  \begin{pmatrix}
    \mathbf{Q}_{\text{\rC\rC}}  & \vline  &     \mathbf{Q}_{\text{\rC\gO}} \\
    \hline
    \mathbf{Q}_{\text{\gO\rC}} &  \vline & \mathbf{Q}_{\text{\gO\gO}}
  \end{pmatrix}
\end{equation}

Although the matrix~$\mathbf{Q}$ has~$(\nO + \nC)^2$ components, for
the diagonal elements~$q_{ii}$ we
have~$q_{ii} = - \sum_{j\neq i} q_{ij}$. Thus, at
most~$(\nO + \nC)(\nO + \nC - 1)$ components of~$\mathbf{Q}$ can be
freely chosen---this coincides with the number of directed edges in a
complete graph with~$\nV=\nO + \nC$ vertices and therefore the maximal
number of rates. 

But if the blocks~$\mathbf{Q}_{\text{\rC\rC}}$
and~$\mathbf{Q}_{\text{\gO\gO}}$ are diagonalisable it is possible to
reparametrise the model~$\mathbf{Q}$ so that all components
within~$\mathbf{Q}_{\text{\rC\rC}}$ and~$\mathbf{Q}_{\text{\gO\gO}}$
except for the diagonal vanish---without changing the dynamics~$\dyn$!
Thus, we consider block-wise similarity transformations~$\mathbf{S}$
introduced by \citet{Kie:89a}

\begin{equation}
  \label{eq:kienker}
  \mathbf{S} = \begin{pmatrix}
    \mathbf{S}_{\text{\rC\rC}}  & \vline  &     \mathbf{0}_{\text{\rC\gO}} \\
    \hline
    \mathbf{0}_{\text{\gO\rC}} &  \vline & \mathbf{S}_{\text{\gO\gO}}
  \end{pmatrix}
\end{equation}

where $\mathbf{S}_{\text{\rC\rC}} \in \mathbb{R}^{\nC \times \nC}$
and~$\mathbf{S}_{\text{\rC\rC}} \in \mathbb{R}^{\nO \times \nO}$ are
invertible matrices and $\mathbf{0}_{\text{\rC\gO}} \in
\mathbb{R}^{\nC \times \nO}$, $\mathbf{0}_{\text{\gO\rC}} \in
\mathbb{R}^{\nO \times \nC}$ are matrices of zeroes. It is easy to see
that

\begin{equation}
  \label{eq:invS}
  \mathbf{S}^{-1} = \begin{pmatrix}
    \mathbf{S}_{\text{\rC\rC}}^{-1}  & \vline  &     \mathbf{0}_{\text{\rC\gO}} \\
    \hline
    \mathbf{0}_{\text{\gO\rC}} &  \vline & \mathbf{S}_{\text{\gO\gO}}^{-1}.
  \end{pmatrix}
\end{equation}

If we consider
models~$\mathbf{Q}'=\mathbf{S}^{-1} \mathbf{Q} \mathbf{S}$, it is
well-known that the
eigenvalues~$-\lambda^1_{\text{\rC}}, \dots,
-\lambda^{\nC}_{\text{\rC}}$
and~$-\lambda^1_{\text{\gO}}, \dots, -\lambda^{\nO}_{\text{\gO}}$ are
invariants but \citet{Kie:89a} demonstrates in his Lemmata 1 and 2
that even the probability distributions~$\fC$, $\fO$, $\fCO$
and~$\fOC$ are preserved under block-wise similarity
transformations. This means that the dynamics~$\dyn$ of a
model~$\mathbf{Q}'=\mathbf{S}^{-1} \mathbf{Q} \mathbf{S}$ is
indistinguishable from the model~$\mathbf{Q}$. For this reason, the
models~$\mathbf{Q}'$ and~$\mathbf{Q}$ are called \emph{equivalent}.


The upper bound~$2\nO \nC$ can now be easily derived if the block
matrices~$\mathbf{Q}_{\text{\rC\rC}}$ and $\mathbf{Q}_{\text{\gO\gO}}$
are diagonalisable. This is true for models that fulfil the
\emph{detailed balance conditions}~\eqref{eq:detbal} which will be
discussed in more detail below. Because the detailed balance
conditions are related to the thermodynamic principle also known as
the Second Law of Thermodynamics that entropy can never decrease in a
closed system (when interpreting a Markov model as a representation of
a chemical system), detailed balance is often assumed for Markov
models of ion channels.

For diagonalisable~$\mathbf{Q}_{\text{\rC\rC}}$ and
$\mathbf{Q}_{\text{\gO\gO}}$, we can find
a 
matrix~$\mathbf{T} = \begin{pmatrix}
  \mathbf{T}_{\text{\rC\rC}}  & \vline  &     \mathbf{0}_{\text{\rC\gO}} \\
  \hline \mathbf{0}_{\text{\gO\rC}} & \vline &
  \mathbf{T}_{\text{\gO\gO}}
\end{pmatrix}$ that brings the blocks~$\mathbf{Q}_{\text{\rC\rC}}$
and~$\mathbf{Q}_{\text{\gO\gO}}$ to diagonal form:

\begin{align}
  \mathbf{Q}' = \mathbf{T}^{-1} \mathbf{Q} \mathbf{T} & = \begin{pmatrix}
    \mathbf{T}_{\text{\rC\rC}}^{-1} \mathbf{Q}_{\text{\rC\rC}}
    \mathbf{T}_{\text{\rC\rC}} & \vline  &
    \mathbf{T}_{\text{\rC\rC}}^{-1} \mathbf{Q}_{\text{\rC\gO}}
    \mathbf{T}_{\text{\gO\gO}} \\
    \hline
    \mathbf{T}_{\text{\gO\gO}}^{-1} \mathbf{Q}_{\text{\gO\rC}}
    \mathbf{T}_{\text{\rC\rC}} &  \vline & \mathbf{T}_{\text{\gO\gO}}^{-1}
    \mathbf{Q}_{\text{\gO\gO}} \mathbf{T}_{\text{\gO\gO}}
  \end{pmatrix}\\
  \label{eq:diagQblocks}
&=  \begin{pmatrix}
  \begin{array}{*{4}c}
    -\lamC[1] & 0& \dots&   0\\
    0 & -\lamC[2] & \ddots
                        & \vdots \\
    \vdots & \ddots &\ddots& 0\\
    0 & \dots & 0 & -\lamC[\nC]                                
  \end{array}
    & \vline  &
 \overbrace{  \mathbf{T}_{\text{\rC\rC}}^{-1} \mathbf{Q}_{\text{\rC\gO}}
    \mathbf{T}_{\text{\gO\gO}}}^{\nO} \} \; {\mathsmaller\nC}\\
    \hline
    {\mathsmaller\nO} \,\{ \underbrace{\mathbf{T}_{\text{\gO\gO}}^{-1} \mathbf{Q}_{\text{\gO\rC}}
    \mathbf{T}_{\text{\rC\rC}}}_{\nC} &  \vline & \begin{array}{*{4}c}
    -\lamO[1] & 0& \dots&   0\\
    0 & -\lamO[2] & \ddots
                        & \vdots \\
    \vdots & \ddots &\ddots& 0\\
    0 & \dots & 0 & -\lamO[\nO]                                
  \end{array}
  \end{pmatrix}
\end{align}

From~\eqref{eq:diagQblocks} it is now clear that the transformed
matrix can at most depend on~$2\nO \nC$ free parameters---the diagonal
elements~$q'_{ii}$ can again be disregarded because they are
determined by the off-diagonal elements and the
matrices~$\mathbf{Q}'_{\text{\rC\gO}}$
and~$\mathbf{Q}'_{\text{\gO\rC}}$ have~$\nO \nC$ parameters each.

This calculation also shows that it is possible to find a
reparametrisation of any fully connected aggregated Markov model to an
aggregated Markov model with~$2\nO\nC$ rate constants (which therefore
fulfils the necessary condition for identifiability). However, as we
will see in Section~\ref{sec:Q3toOCO} where this transformation is
carried out explicitly for the fully connected three-state model, this
does not always yield a feasible model---as we will observe, there are
parameter sets for which some of the rates in the reparametrised model
are negative.

The matrix~\eqref{eq:diagQblocks} defines a model structure where all
transition rates from open to other open states or between closed
states vanish---all transitions occur between open and closed
states. This structure has been proposed as a \emph{canonical from}
for models with a given number of~$\nO$ open and~$\nC$ closed
states. Canonical forms provide models that are parameter-identifiable
and can be used as representatives for equivalence classes of
models. The particular form~\eqref{eq:diagQblocks} has been studied by
\citet{Bau:87a,Kie:89a} and is referred to as \emph{Bauer-Kienker uncoupled
(BKU) form} in \citet{Bru:05a}. For a brief discussion of canonical
forms see the Discussion (Section~\ref{sec:discussion}).



  


\subsection{Interpretation of sojourn time distributions}
\label{sec:sojournModel}

\begin{figure}[htbp]
  \centering
  \subfloat[$\fO(\tO)$]{%
        \label{fig:fO}
      \includegraphics[width=0.45\linewidth]{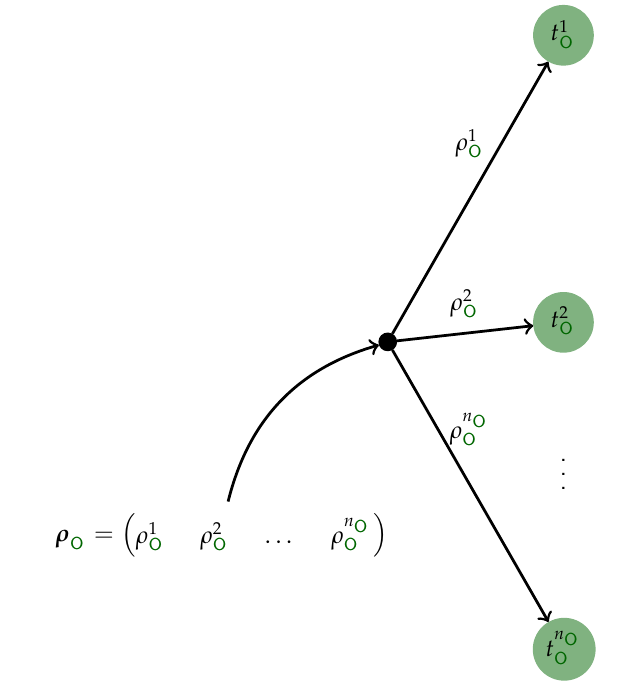}%
    }%
    \quad
    \subfloat[$\fOC(\tO, \tC)$]{%
    \label{fig:fOC}
    \includegraphics[width=0.5\linewidth]{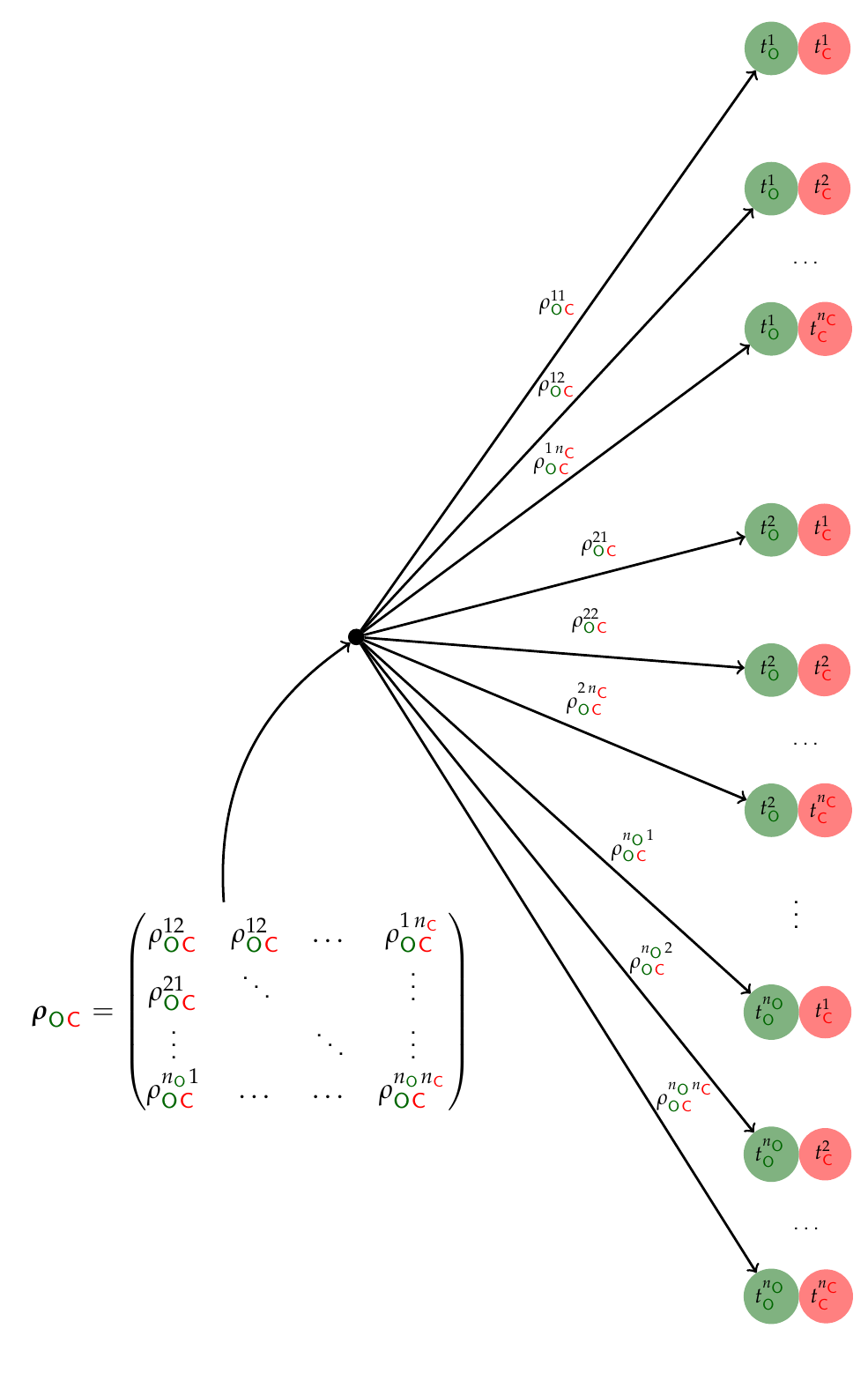}%
    }%
    \caption{Graphical illustration of the sojourn time
      distribution 
      $\fO(\tO)$ and the bivariate distribution~$\fOC(\tO, \tC)$ as
      mixtures of exponential distributions---the
      distributions~$\fC(\tO)$ and~$\fCO(\tC, \tO)$ are analogous. It
      shows that sojourn times~$\tO$ are generated in a two-stage
      process. First, from the finite distribution defined
      by~$\boldsymbol{\rho}_{\text{\gO}}$, an exponential
      distribution~$\tO[i]$ is chosen which then generates an open
      time~$\tO$, see (a). Similarly, the
      matrix~$\boldsymbol{\rho}_{\text{\gO\rC}}$ defines a finite
      distribution over the pairs of exponential distributions. After
      choosing a pair
      $(\tO[i],\tC[j])$, $i=1, \dots, \nO$, $j=1, \dots, \nC$, it
      generates an open time~$\tO$ and a subsequent closed time~$\tC$,
      see (b).}
  \label{fig:sojourndist}
\end{figure}

The reduction 
of aggregated Markov models to the bivariate distributions~$\fOC$
and~$\fCO$ \citep{Fre:85a, Fre:86a} leads to an 
alternative interpretation of the dynamics~$\dyn$. 
This leads to a more abstract view 
compared to the illustrative transitions between open and closed
states which are assigned additional biophysical significance by
associating them with different conformational states of the channel
protein. Figure~\ref{fig:sojourndist} illustrates that the
distributions~$\fC$, $\fO$, $\fOC$ and~$\fCO$ can be interpreted as
two-stage processes where a finite distribution is used for
determining 
which distributions are used for generating either sojourn times or
pairs of subsequent sojourn times. Figure~\ref{fig:fO} illustrates how
the coefficients~$\boldsymbol{\rho}_{\text{\gO}}$ define a finite
distribution that stochastically selects if an open time is generated
from a ``slow'' or ``fast'' exponential state. As shown in
Figure~\ref{fig:fOC} this is extended to the distribution~$\fOC$ which
generates an open time followed by a closed time.  After an open and a
closed exponential state have been chosen by the finite distribution
defined by the components~$\rho_{\text{\gO\rC}}^{ij}$ of the
matrix~$\boldsymbol{\rho}_{\text{\gO\rC}}$, a pair~$(\tO, \tC)$ of
subsequent open and closed times is generated.

\subsection{\emph{Example:} The fully connected three-state model}
\label{sec:nonidExample}

In order to illustrate the consequences of this abstract result by
\citet{Fre:85a,Fre:86a} with a concrete example, let us consider the
simplest case of a non-identifiable model, the fully connected
three-state model. We follow and expand the presentation
from~\citet{Kie:89a}:

\begin{equation}
  \label{eq:Q3full}
  Q =
  \begin{pmatrix}
    -q_{12} - q_{13} & q_{12} & \vline & q_{13}\\
    q_{21} & -q_{21} -  q_{23} & \vline &  q_{23}\\
    \hline
    q_{31} &q_{32} &\vline & -q_{32} -  q_{31}
  \end{pmatrix}
\end{equation}

The matrix~$Q$ in~\eqref{eq:Q3full} is partitioned in the closed
compartment~$\text{\rC}=\{ \text{\rC[1], \rC[2]} \}$ 
and the open compartment~$\text{\gO}=\{ \text{\gO[3]} \}$
. It is helpful to summarise the parameters of the model via the sojourn time
distributions~$\fO(t)$ and~$\fC(t)$:

\begin{align}
  \label{eq:Q3fullfC}
  \fC(\tC) & = \rho \lamC[1]\exp (-\lamC[1] \tC ) +
  (1-\rho) \lamC[2] \exp (-\lamC[2] \tC ), \\
    \label{eq:Q3fullfO}
  \fO(\tO) & = \lamO[1] \exp (-\lamO[1] \tO )
\end{align}

The two distributions~$\fC$ and $\fO$ together depend on four
independent parameters---the weight~$\rho$ of the first component in
the mix of two exponential distributions that~$\fC$ consists of as
well as the rates~$\lamC[1]$, and
$\lamC[2]$ and~$\lamO[1]$ which are
eigenvalues of submatrices of the matrix~$Q$. 

Because eigenvalues depend in an increasingly complex way on the
components of a matrix as the dimension increases, it is advantageous
to alternatively consider

\begin{align}
  \label{eq:trQc3}
  \Tc &:= \Tr Q_{\text{\rC \rC}} = -q_{12} - q_{13} - q_{21} - q_{23} =
                           -\lamC[1] -  \lamC[2],\\
    \label{eq:detQc3}
  D_{\text{\rC}} &:= \Det Q_{\text{\rC \rC}} = q_{12} q_{23} + q_{13} q_{21} + q_{13}
                   q_{23} =  \lamC[1] \cdot \lamC[2],\\
  \label{eq:SQ3}
  S & :=  Q_{\text{\gO \rC}}  Q_{\text{\rC \gO}} = q_{13} q_{31} + q_{23} q_{32} , \\
  \label{eq:trQo3}
      T_{\text{\gO}} & := \Tr  Q_{\text{\gO}} = -q_{31} - q_{32} = -\lamO[1].
\end{align}

when evaluating transformations of the generator~$Q$. In order to
investigate non-identifiability, we are interested in
reparametrisations of the matrix~$Q$ that alter the rate
constants~$q_{ij}$ but leave the parameters of the sojourn time
distributions invariant.

For the three-state model, the rates of the closed-time
distribution~$\fC(\tC)$ are obtained as solutions of the
characteristic polynomial of the matrix~$Q_{\text{\rC \rC}}$
\begin{equation}
  \label{eq:Q3charpol}
  \lambda^2 -  T_{\text{\rC}} \lambda + D_{\text{\rC}} = 0
  \end{equation}

  The weight~$\rho$ can be found by elementary but relatively tedious
  calculations

  \begin{equation}
    \label{eq:Q3rho}
    \rho = \frac{S
      -\lamO[1] \lamC[2]}{\lamO[1] \cdot
      \left( \lamC[1] - \lamC[2] \right)} =
    \frac{\frac{S}{\lamO[1]} -\lamC[2]}{\lamC[1] - \lamC[2]}.
  \end{equation}

  Because, in order to leave the closed time distribution~$\fC(\tC)$
  unchanged, the parameter~$\rho$ needs to be preserved by a
  reparametrisation and the expression~\eqref{eq:Q3rho} shows that~$S$
  must be an invariant when reparametrising the model because all
  other parameters on the right-hand side of~\eqref{eq:Q3rho} are
  invariants. It also shows that it is possible to parameterise the
  three-state model, so that~$\rho<0$. In this case, the closed-time
  distribution~$\fC$ is a signed mixture of exponential distributions,
  rather than an ordinary mixture distribution for which all
  coefficients must be positive.

  We will see in the following Section~\ref{sec:Q3toOCO} that
  for~$\rho>0$ where~$\fC$ is a mixture of exponential distributions,
  the fully connected three-state model can be transformed to a
  reduced model with fewer rate constants, decreasing the original $6$
  parameters to model with $4$ parameters. In contrast, this will not
  be possible if~$\fC$ is a signed mixture of exponential
  distributions.

\subsubsection{Transformation of the general three-state model to the
  model~\rC\gO\rC}
\label{sec:Q3toOCO}

We will attempt to reparametrise the model~$Q$ to obtain a
model~$\tilde{Q}$ with the same sojourn time distributions~$\fC$
and~$\fO$. The model \rC\gO\rC\ is a special case of the three-state
model where both closed states~$C_1$ and~$C_2$ are adjacent to the
open state~$O_3$ but direct transitions between the closed states are
not possible---$\tilde{q}_{12}=\tilde{q}_{21}=0$. Because the
matrix~$\tilde{Q}_{\text{\rC \rC}}=
\begin{pmatrix}
  -\tilde{q}_{13} & 0\\
  0 & -\tilde{q}_{23}
\end{pmatrix}
$ is diagonal it follows that the rate constants~$\tilde{q}_{13}$
and~$\tilde{q}_{23}$ must coincide with the
eigenvalues~$\lamC[1]$ and~$\lamC[2]$. For
the remaining rate constants, in order to keep~$\To$
and~$S$ invariant, we obtain linear equations
in~$\tilde{q}_{31}$ and~$\tilde{q}_{32}$:

\begin{align}
  \label{eq:3}
  \tilde{q}_{31} + \tilde{q}_{32} & = \lambda^1_{\text{\gO}}
  \\
  \tilde{q}_{13} \tilde{q}_{31} + \tilde{q}_{23}  \tilde{q}_{32} =
  \lambda^1_{\text{\rC}} \tilde{q}_{31} +\lambda^2_{\text{\rC}}
  \tilde{q}_{32} &= S
\end{align}


Interestingly, the solution

\begin{equation}
  \label{eq:q31OCO}
  \tilde{q}_{31} =
  \frac{S-\lamO[1] \lamC[2]}{\lamC[1] - \lamC[2]}= \lamO[1]
    \frac{\frac{S}{\lamO[1]}
      -\lamC[2]}{\lamC[1] - \lamC[2]} = \lamO[1] \cdot \rho
  \end{equation}

  so that the model~\rC\gO\rC\ can be easily obtained from the
  distributions~$\fO$ and~$\fC$ by assigning the parameters $\rho$,
  $\lamC[1]$, $\lamC[2]$ and~$\lamO[1]$ to the appropriate rate
  constants of the model \rC\gO\rC\ (this is, of course, the specific
  example of the transformation used in Section~\ref{sec:altproof} to
  derive the upper bound~$2\nO\nC$ for the maximum number of rates for
  the model with~$\nV$ vertices):

\begin{equation}
  \label{eq:Q3OCO}
  \tilde{Q} =
  \begin{pmatrix}
    -\lamC[1] &0 & \vline & \lamC[1]\\
    0 & -\lamC[2] & \vline &  \lamC[2]\\
    \hline
    \rho \cdot \lamO[1]  &     (1-\rho) \cdot \lamO[1] &\vline & -\lamO[1] 
  \end{pmatrix}
\end{equation}

It seems that the non-identifiability problem---at least for the
simple case of the full three-state model---has a very simple
solution (deceptively simple as it turns out). Instead of considering the complete parameter space of the
fully connected three-state model, the model is reduced to an
equivalent \rC\gO\rC\ model.

But this overlooks that this reduction is not always possible in a
meaningful way---for some possible parameter choices of three-state
model, a negative value for~$\rho$ is obtained. In this case, the rate
constant~$\tilde{q}_{31}$ is negative which contradicts the underlying
assumption that the rate constants of a Markov generator are all
non-negative.


\subsubsection{Continuous reparametrisation of the fully connected
  three-state model}
\label{sec:contequiv}

The fully connected model cannot only be transformed to the
model~\rC\gO\rC\ but, in fact, to a continuous set of models
parametrised by two of the rates. By considering that the quantities in
\eqref{eq:trQc3}-\eqref{eq:trQo3} need to be invariant to preserve the
sojourn time distributions~$\fO$ and~$\fC$ these quantities can be
taken as invariants that different parametrisations need to fulfil. We
can, for example, choose~$q_{13}$ and~$q_{23}$ as free parameters and
express the remaining rate constants as functions of~$q_{13}$
and~$q_{23}$ for given~$\To, \Tc, \Dc$ and~$S$.

From the relationship

\begin{equation}
  \label{eq:4}
  q_{23}^2 + T_{\text{\rC}} q_{23} + D_{\text{\rC}} = q_{21}(q_{13} -
  q_{23} ) 
\end{equation}

we can calculate:

\begin{equation}
  \label{eq:q21Tq13q23}
  q_{21} = \frac{q_{23}^2 + T_{\text{\rC}} q_{23} + D_{\text{\rC}}}
  {q_{13} - q_{23}}
\end{equation}

This expression for~$q_{21}$ allows us---using the trace~$Tc$ in~\eqref{eq:trQc3}--- to derive

\begin{equation}
  \label{eq:q12Tq13q23}
  q_{12} =- \frac{q_{13}^2 + T_{\text{\rC}} q_{13} + D_{\text{\rC}}}
  {q_{13} - q_{23}}
\end{equation}

Similarly, with $S$ from~\eqref{eq:SQ3} after replacing~$q_{32}$
using the trace~$\To$, equation~\eqref{eq:trQo3}, we can calculate:

\begin{equation}
  \label{eq:q31Tq13q23}
  q_{31} = \frac{S + T_{\text{\gO}} q_{23}} {q_{13} - q_{23}}
\end{equation}

and finally
\begin{equation}
  \label{eq:q32Tq13q23}
  q_{32} = - T_{\text{\gO}} - \frac{S + T_{\text{\gO}} q_{23}} {q_{13}
    - q_{23}} = -\frac{S + T_{\text{\gO}} q_{13}} {q_{13}
    - q_{23}}
\end{equation}

\subsubsection{Transformation of the general three-state model to the
  model~\rC\rC\gO}
\label{sec:Q3toOCC}

Setting~$q_{13}=0$ in the
equations~\eqref{eq:q21Tq13q23}-\eqref{eq:q32Tq13q23} we can calculate
the rates of a model where~$q_{13}$ vanishes:

\begin{equation}
  \label{eq:Q3CCO}
  \tilde{Q} =
  \begin{pmatrix}
    -\frac{\lamC[1]\lamC[2]}{q_{23}} & \frac{\lamC[1] \lamC[2]}{q_{23}} & \vline & 0\\
    -\frac{ \left(q_{23}  - \lamC[1] \right)\left(q_{23}  - \lamC[2] \right)}{q_{23}} & \frac{-\left(\lamC[1] +
      \lamC[2]\right)q_{23} + \lamC[1] \lamC[2]}{q_{23}}& \vline &  q_{23}\\
    \hline
   \lamO[1] -\frac{S}{q_{23}} &    \frac{S}{q_{23}}&\vline & -\lamO[1]
  \end{pmatrix}
\end{equation}

Interestingly, $q_{13}=0$ does not imply~$q_{31}=0$. But~$q_{31}$ does
vanish if we require the \emph{detailed balance conditions}. By
definition, the detailed balance conditions are

\begin{align}
  \label{eq:detbal}
  \pi^i_{\text{\rC}} q_{i3} &=   \pi^3_{\text{\gO}} q_{3i}, \quad
                              i=1,2\\  
  \pi^3_{\text{\gO}} q_{3i} &=   \pi^i_{\text{\rC}} q_{i3} \\
  \pi^1_{\text{\rC}} q_{12} &=   \pi^2_{\text{\rC}} q_{21}.
\end{align}

If a vector~$\boldsymbol{\pi}$ exists that fulfils these conditions it
is a \emph{stationary distribution} of the Markov model. It can be
shown that these conditions are fulfilled provided that

\begin{equation}
  \label{eq:detbalQ3}
  q_{31} q_{12} q_{23}=  q_{13} q_{32} q_{21},
\end{equation}

in general, the products of rates along all cycles that involve three
or more states must be equal in either direction.  With detailed
balance it, of course, follows that if $q_{13}=0$,
{$q_{31}=\lamO[1] - \frac{S}{q_{23}}$} must vanish as
well{---if $q_{13}=0$ the left-hand side vanishes
  of~\eqref{eq:detbalQ3} which implies that one of the rates on the
  left-hand side must be zero as well}. Thus,
$q_{23}=\frac{S}{\lamO[1]}$ and we have

\begin{equation}
  \label{eq:Q3CCOdetBal}
  \tilde{Q} =
  \begin{pmatrix}
    -\frac{\lamO[1] \lamC[1] \lamC[2]}{S} &  \frac{\lamO[1] \lamC[1] \lamC[2]}{S} & \vline & 0\\
    - \frac{\lamO[1]}{S} \left(\frac{S}{\lamO[1]} - \lamC[1]
    \right)\left(\frac{S}{\lamO[1]} - \lamC[2] \right) &
    \frac{-(\lamC[1] + \lamC[2])S + \lamO[1] \lamC[1] \lamC[2]}{S}& \vline &
    \frac{S}{\lamO[1]}\\
    \hline 0 & \lamO[1] &\vline & -\lamO[1]
  \end{pmatrix}.
\end{equation}

It is also possible to find a version of the model~\gO\rC\rC\ for
which~$q_{12}=0$ but~$q_{21} \neq 0$ (this model does
therefore not fulfil detailed balance). Substituting~$q_{13}=\lamC[1]$
in~\eqref{eq:q21Tq13q23} and~\eqref{eq:q31Tq13q23} and
rewriting~$q_{31}$ and $q_{32}$ using the expression for~$\rho$
from~\eqref{eq:Q3rho} yields

\begin{equation}
  \label{eq:Q3OCOnobal}
  \tilde{Q} =
  \begin{pmatrix}
    -\lamC[1] &0 & \vline & \lamC[1]\\
    \lamC[2] - q_{23} & -\lamC[2] & \vline &  q_{23} \\
    \hline
    \rho \left( 1 - \frac{1-\rho}{\rho} \frac{\lamC[2] -
        q_{23}}{q_{23} - \lamC[1]} \right) \cdot \lamO[1]  &
    \left[1-\rho \left( 1 - \frac{1-\rho}{\rho} \frac{\lamC[2] -
          q_{23}}{q_{23} - \lamC[1]} \right) \right] \cdot \lamO[1] &\vline & -\lamO[1] 
  \end{pmatrix}
\end{equation}

For the resulting model in~\eqref{eq:Q3OCOnobal} it shows that
for~$q_{12}=0$, an infinite set of equivalent models can be obtained
by continuously varying~$q_{23} \in \left(\lamC[1], \lamC[2] \right]$.

\begin{figure}[h!]
  \centering
  \subfloat[Reducible to acyclic model]{%
    \label{fig:Q3red}
    \includegraphics[width=0.45\linewidth]{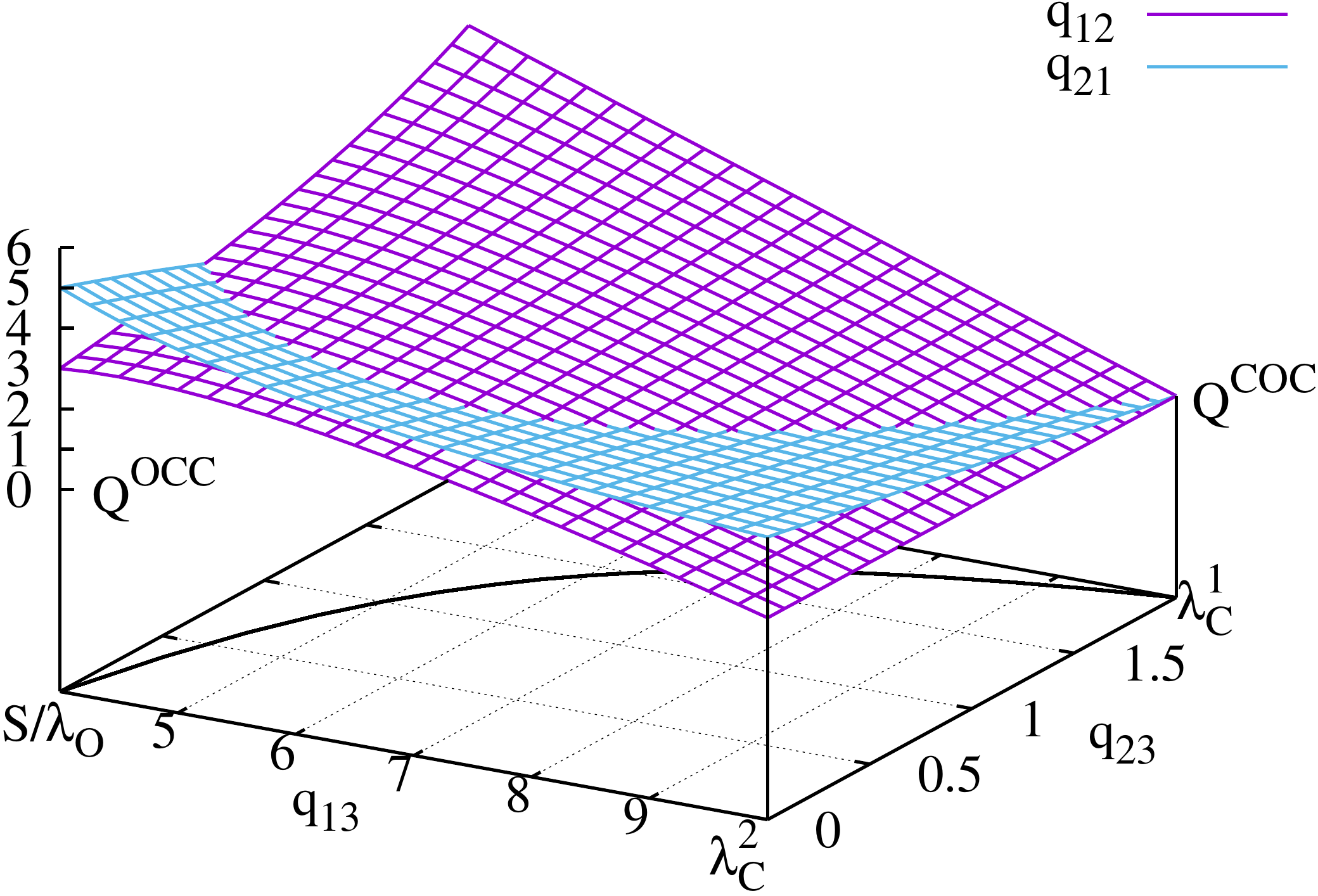}%
  \;
    \includegraphics[width=0.45\linewidth]{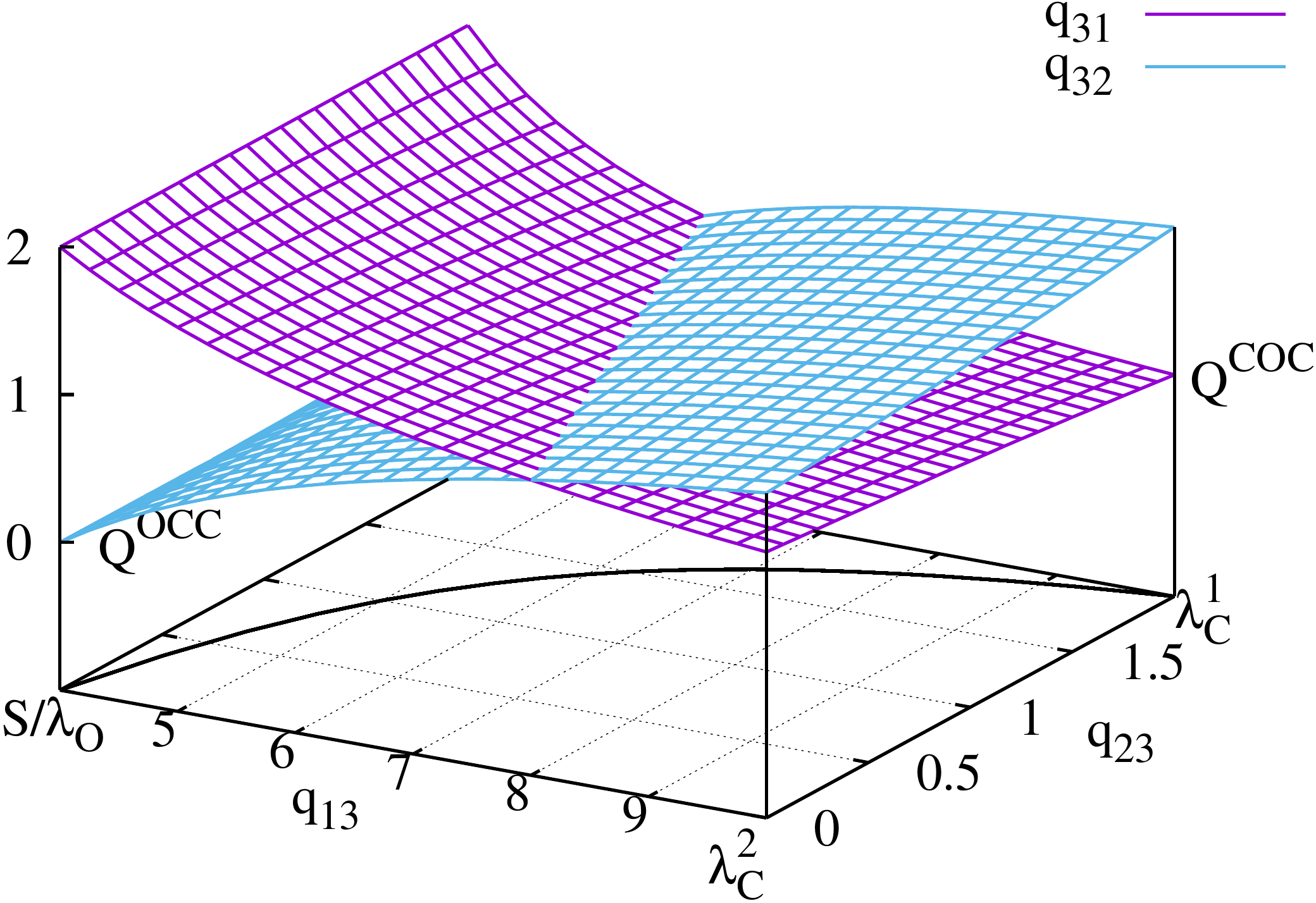}%
  }%
  \\
  \subfloat[Not reducible to acyclic model]{%
    \label{fig:Q3irred}
    \includegraphics[width=0.45\linewidth]{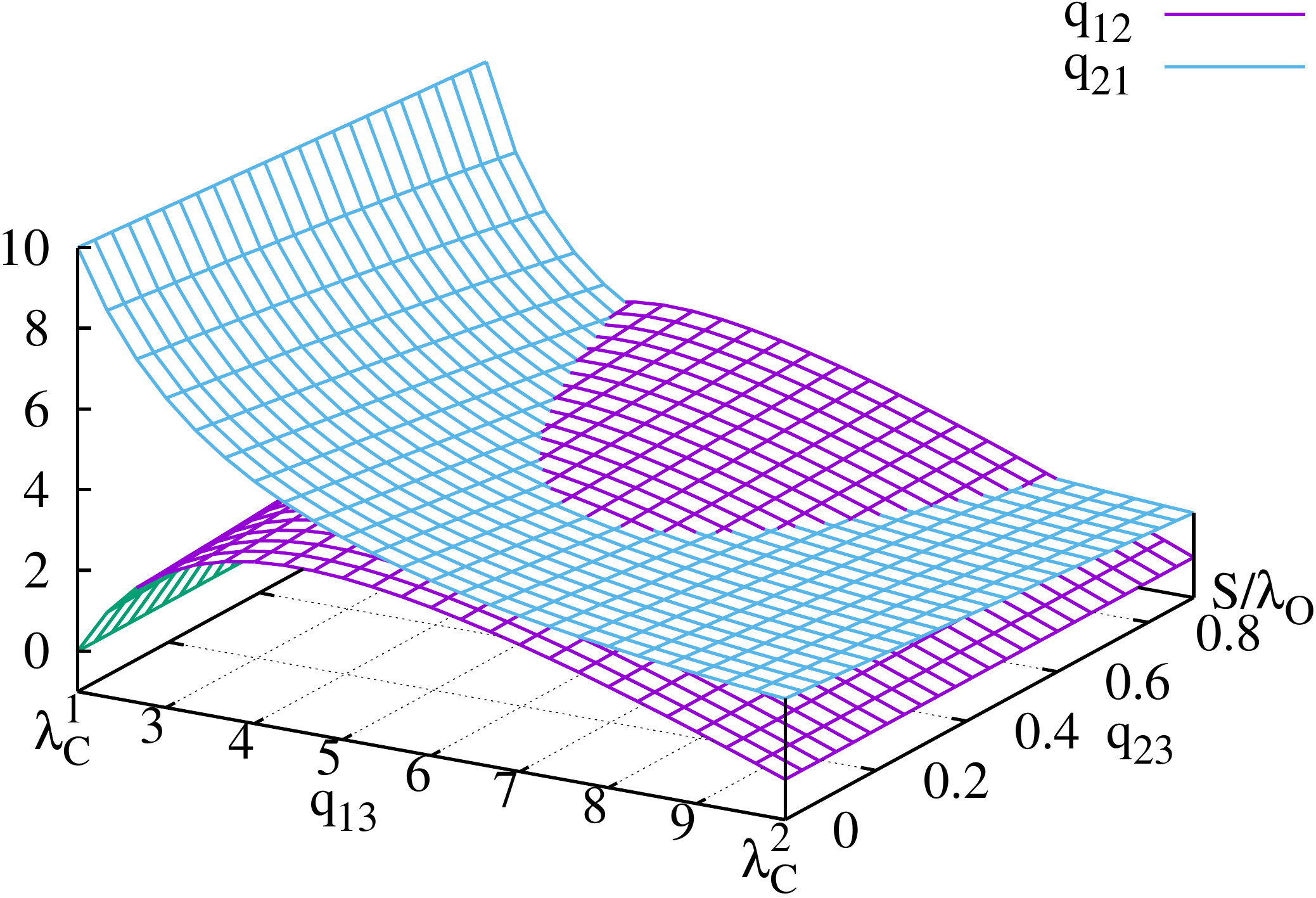}%
  \;
    \includegraphics[width=0.45\linewidth]{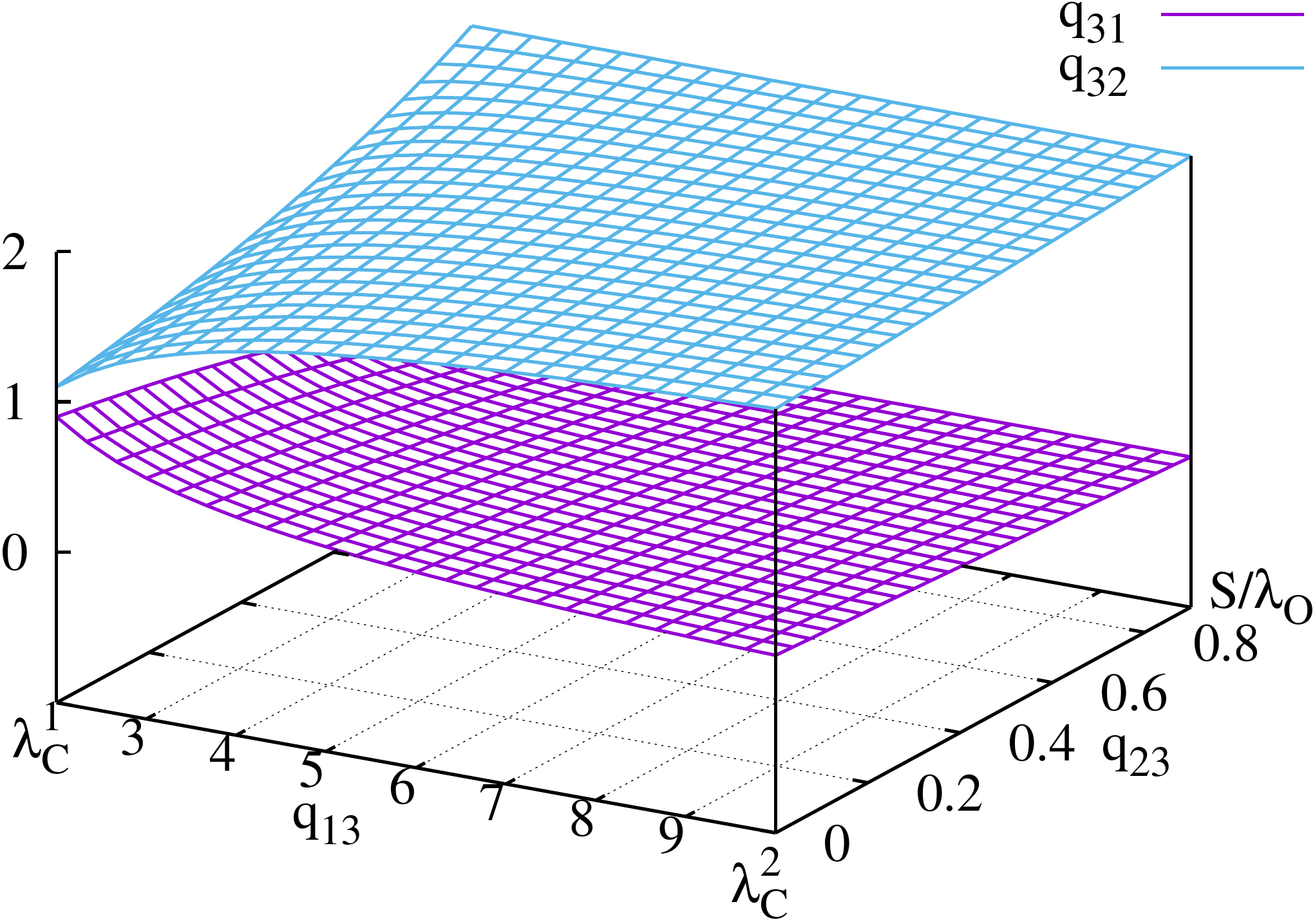}%
  }%
  \caption{Summary of the analysis of the complete three-state
    model. The rates~$q_{13}$ and~$q_{23}$ connecting the closed
    states~$C_1$ and $C_2$ with the open state~$O_3$. Without loss of
    generality we assume~$q_{13}>q_{23}$ and~$\lamC[1] < \lamC[2]$. In
    (a), $\frac{S}{\lamO[1]} \in \left[\lamC[1], \lamC[2] \right]$. In
    this case, for all
    choices~$q_{13} \in \left[\frac{S}{\lamO[1]}, \lamC[2] \right]$
    and~$q_{23} \in \left[0, \lamC[1]\right]$, parametrisations can be
    found which produce the same stochastic dynamics. The parameter
    sets obtained for combinations of~$q_{13}$ and~$q_{23}$ along the
    curve on the base of the plot fulfil detailed balance,
    see~\eqref{eq:detbalQ3}. For~$q_{13}=\frac{S}{\lamO[1]}$,
    $q_{23}=0$ we obtain the model~\gO\rC\rC, for~$q_{13}=\lamC[2]$,
    $q_{23}=\lamC[1]$ we obtain the model~\rC\gO\rC. The situation is
    fundamentally different in (b)
    where~$\frac{S}{\lamO[1]} < \lamC[1]$. Here, the fully connected
    three-state model cannot be reduced to the acyclic
    model~\gO\rC\rC\ or~\rC\gO\rC\ with non-negative rates. Also, in
    this case the closed time distribution~$\fC$ is not a mixture but
    a signed mixture of exponentials which generates a non-monotonous
    distribution. The set of equivalent models with non-negative rates
    in this case extends over the
    intervals~$q_{13} \in \left[ \lamC[1], 
      \lamC[2] \right]$
    and~$q_{23} \in \left[ 0, \frac{S}{\lamO[1]} \right]$. The detailed
    balance condition~\eqref{eq:detbalQ3} cannot be fulfilled for any
    of these models.}
  \label{fig:Q3}
\end{figure}

\subsubsection{Summary}
\label{sec:summary}

Our analysis of the fully connected three-state model 
has illustrated the two aspects of identifiability introduced in the
introduction--- parameter identifiability and the identifiability
of model structure.

Our analysis shows that, at least for the example of the fully
connected three-state model, \emph{parameter identifiability} and
\emph{identifiability of model structure} are closely linked as
visualised in Figure~\ref{fig:Q3}. Both Figures~\ref{fig:Q3red}
and~\ref{fig:Q3irred} show parameter sets~$q_{ij}$ of the
fully-connected three-state model 
that preserve the four invariants~$\Tc$, $\Dc$, $\To$ and~$S$,
see~\eqref{eq:trQc3}-\eqref{eq:trQo3}. Because the three-state model
is parametrised by six rates, we obtain an equivalence class that
depends on two parameters such as the rates~$q_{13}$ and~$q_{23}$, for
example, as in Section~\ref{sec:contequiv}. Both rates are constrained
to finite intervals in order to ensure that all six rate constants of
the three-state model are non-negative.

When considering this equivalence class of models, two fundamentally
different cases can be distinguished---without loss of generality, we
assume that~$\lamC[1]<\lamC[2]$:

\begin{enumerate}
\item If~$\frac{S}{\lamO[1]} \in [\lamC[1], \lamC[2] ]$, the
  models~\rC\gO\rC\ and~\rC\rC\gO\ are contained in the equivalence
  class. In this case, parameter identifiability can be resolved by
  selecting either~\rC\gO\rC\ or~\rC\rC\gO\ as a representative of the
  equivalence class. In this way, non-identifiable models with six
  rates can be \emph{reduced} to either of the two identifiable
  models~\rC\gO\rC\ or~\rC\rC\gO\ with four rates. This case is
  visualised in Figure~\ref{fig:Q3red}.
\item If~$\frac{S}{\lamO[1]} \not\in [\lamC[1], \lamC[2] ]$, it is not
  possible to reduce models in the equivalence class to
  either~\rC\gO\rC\ or~\rC\rC\gO\ so that the rates are all
  non-negative. Thus, in this case, the parameter non-identifiability
  cannot be resolved because it is not possible to choose any of the
  parameter-identifiable models~\rC\gO\rC\ or~\rC\rC\gO\ as a
  representative. Also, in this case, the detailed balance
  conditions~\eqref{eq:detbal} cannot be fulfilled. Parameter sets of
  this type are visualised in Figure~\ref{fig:Q3irred}.
\end{enumerate}

By considering~\eqref{eq:Q3rho}, we find that the
condition~$\frac{S}{\lamO[1]} \in [\lamC[1], \lamC[2] ]$ is equivalent
to~$\rho>0$. This means that models are reducible to an identifiable
model if the closed time distribution~$\fC$ is a mixture of
exponentials. In the opposite
case~$\frac{S}{\lamO[1]} \not\in [\lamC[1], \lamC[2] ]$ we find, in
contrast, that the closed time distribution is a signed mixture of
exponential distribution---here, the coefficients~$\rho$ and~$1-\rho$
of the closed time distribution~$\fC$ sum to one but either~$\rho$
or~$1-\rho$ is negative.

Because in this case, where the closed time distribution~$\fC$ is a
non-monotonous signed mixture of exponentials, it is not possible to
find a representative that is parameter-identifiable, this means that
there are dynamics~$\dyn$ which cannot be represented by an
identifiable model which means that it is not possible to identify an
unambiguous ``mechanism'' that can be represented using aggregated
Markov models which generates this dynamics~$\dyn$. To the best of our
knowledge this phenomenon which deserves further attention has not
been systematically investigated for more aggregated Markov models
with a larger number of states.

\subsection{Model structure and biophysical
  mechanism\protect\footnote{
    More than 10 years ago, Edmund Crampin and I had several
    passionate (but enjoyable) discussions about the ``right''
    approach for modelling ion channels. Edmund suggested that I look
    at the relationship between different models for ligand binding
    and non-identifiability. This section is a first attempt to
    investigate the implications of non-identifiability on mechanistic
    models.}%
}
\label{sec:biophysics}

We will now investigate the question if natural mechanistic models for
the dependency of a ligand-gated ion channel on the ligand
concentration~$c$ can help us decide which of the models~\rC\gO\rC\
or~\rC\rC\gO\ might be preferable for a given data set. Although we
have shown previously that both models are equivalent, they represent
quite different mechanisms if we assume that the transitions between
the model states are modulated by ligand binding sites. The
model~\rC\gO\rC\ represents two \emph{independent ligand binding}
sites which can either facilitate or inhibit the transition to the
open state. In contrast, \rC\rC\gO\ models \emph{sequential
  binding}---a first ligand binding site increases the transition from
the closed state~\rC[1] to \rC[2] whereas a second binding site (which
becomes accessible when the channel has transitioned to~\rC[2]) then
activates the channel. To transform between the two models we can
use~\eqref{eq:Q3OCO} and~\eqref{eq:Q3CCO} to directly relate the rates
of the models~\rC\gO\rC\ and~\rC\rC\gO: 

\begin{align}
  \label{eq:q12CCO}
  q_{12}^{\text{\rC\rC\gO}} & =
                              \frac{q_{13}^{\text{\rC\gO\rC}}q_{23}^{\text{\rC\gO\rC}}\left(
                              q_{31}^{\text{\rC\gO\rC}} + q_{32}^{\text{\rC\gO\rC}}
                              \right)}{q_{13}^{\text{\rC\gO\rC}}
                              q_{31}^{\text{\rC\gO\rC}} +
                              q_{23}^{\text{\rC\gO\rC}}
                              q_{32}^{\text{\rC\gO\rC}}} \\
    \label{eq:q21CCO}
    q_{21}^{\text{\rC\rC\gO}} & =
                              \frac{ \left(q_{13}^{\text{\rC\gO\rC}} -
                                q_{23}^{\text{\rC\gO\rC}} \right)^2}{q_{13}^{\text{\rC\gO\rC}}
                              q_{31}^{\text{\rC\gO\rC}} +
                              q_{23}^{\text{\rC\gO\rC}}
                              q_{32}^{\text{\rC\gO\rC}}}
                             \frac{q_{31}^{\text{\rC\gO\rC}}q_{32}^{\text{\rC\gO\rC}}}{q_{31}^{\text{\rC\gO\rC}} + q_{32}^{\text{\rC\gO\rC}}} 
  \\
  \label{eq:q23CCO}
   q_{23}^{\text{\rC\rC\gO}} & =
                               \frac {q_{13}^{\text{\rC\gO\rC}}
                              q_{31}^{\text{\rC\gO\rC}} +
                              q_{23}^{\text{\rC\gO\rC}}
                              q_{32}^{\text{\rC\gO\rC}}}
                               {
                               q_{31}^{\text{\rC\gO\rC}} + q_{32}^{\text{\rC\gO\rC}}
                               } \\
  \label{eq:q32CCO}
   q_{32}^{\text{\rC\rC\gO}} &= q_{31}^{\text{\rC\gO\rC}} + q_{32}^{\text{\rC\gO\rC}}
\end{align}

One aspect of~\eqref{eq:q12CCO}-\eqref{eq:q32CCO} is that these
expressions clearly show that an arbitrary model of the
from~\rC\gO\rC\ can be transformed to a~\rC\rC\gO\ model---if the
rates $q_{13}^{\text{\rC\gO\rC}}$, $q_{23}^{\text{\rC\gO\rC}}$,
$q_{31}^{\text{\rC\gO\rC}}$ and $q_{32}^{\text{\rC\gO\rC}}$ are
positive, the rates $q_{12}^{\text{\rC\rC\gO}}$,
$q_{21}^{\text{\rC\rC\gO}}$, $q_{23}^{\text{\rC\rC\gO}}$ and
$q_{32}^{\text{\rC\rC\gO}}$ on $q_{13}^{\text{\rC\gO\rC}}$,
$q_{23}^{\text{\rC\gO\rC}}$, $q_{31}^{\text{\rC\gO\rC}}$ and
$q_{32}^{\text{\rC\gO\rC}}$ of the transformed model are positive as
well.

But our primary interest is to investigate the effect of this
transformation for different assumptions on the ligand dependencies of
the model~\rC\gO\rC. For simplicity, we will assume that all
dependencies are modelled via mass action
. We will drop the superscripts for the rates of the models~\rC\rC\gO\
and~\rC\gO\rC, instead we will use~$q_{13}, q_{23}, q_{31}$ and
$q_{32}$ when we refer to rates of the model~\rC\gO\rC\
and~$\tilde{q}_{12}, \tilde{q}_{13}, \tilde{q}_{21}$ and
$\tilde{q}_{31}$ when we refer to rates in the model~\rC\rC\gO\
obtained via transformation of the model~\rC\gO\rC.

\subsubsection{Two activating binding site}
\label{sec:Q3ligandtwoact}

First, we assume that the channel has two activating binding
sites. This implies that the rates~$q_{13}$ and~$q_{23}$ entering the
open state are ligand-dependent and the rates~$q_{31}$ and~$q_{32}$
are constant i.e. independent from the ligand concentration~$c$:

\begin{align}
  \label{eq:Q3twoact}
  q_{13} = k_{13} c,   q_{23} = k_{23} c,   q_{31} = k_{31},   q_{32} = k_{32}.
\end{align}

Replacing \eqref{eq:q12CCO}-\eqref{eq:q32CCO} we observe that the
resulting rate constants for the model~\rC\rC\gO\ are:

\begin{align}
  \label{eq:Q3twoacttrans}
  \tilde{q}_{12} = \tilde{k}_{12} c,     \tilde{q}_{21} = \tilde{k}_{21} c,     \tilde{q}_{23} =
  \tilde{k}_{23} c,     \tilde{q}_{32} = \tilde{k}_{32}.
\end{align}

with, as can simply be seen,

\begin{align}
  \label{eq:Q3twoacttranspar}
  \tilde{k}_{12} & = \frac{k_{13} k_{23} (k_{31} + k_{32}) }{k_{13}
                   k_{31} +
                   k_{23} k_{32}} \\
  \tilde{k}_{21} & =  \frac{ \left(k_{13} - k_{23} \right)^2}{k_{13}
                   k_{31} + k_{23} k_{32}}
                   \frac{k_{31} k_{32}}{k_{31}
                   + k_{32}} \\
   \tilde{k}_{23} & =\frac {k_{13} k_{31} + k_{23} k_{32}}
                               {
                               q_{31}^{\text{\rC\gO\rC}} + q_{32}^{\text{\rC\gO\rC}}
                               } \\
   \tilde{k}_{32} &= k_{31} + k_{32}
\end{align}

From a modelling point of view, the resulting \rC\rC\gO\ model is
unusual. It appears that the activating binding sites that facilitate
the transitions from the closed states~\rC[1] and~\rC[2] to the open
state~\gO[3] translate to activating binding sites that enable the
transition from~\rC[1] via~\rC[2] to~\gO[3]. But although unbinding
from binding sites would normally assumed to be ligand-independent,
the transition~$q_{21}$ from~\rC[2] to \rC[1] is again a mass action
rate dependent on~$c$. This implies that in order to reproduce the
behaviour of the \rC\gO\rC\ model, a ligand-dependent rate that
inhibits the channel for large~$c$ by transitioning to the
state~\rC[1]\ seems to be required.

\subsubsection{One activating binding site}
\label{sec:Q3ligandoneact}

Now, we assume that, of the two transitions to the open state, one is
activated by ligand binding whereas the other transition is
ligand-independent:

\begin{align}
  \label{eq:Q3oneact}
  q_{13} = k_{13} c,   q_{23} = k_{23},   q_{31} = k_{31},   q_{32} = k_{32}.
\end{align}

Replacing \eqref{eq:q12CCO}-\eqref{eq:q32CCO} we observe that the
resulting rate constants for the model~\rC\rC\gO\ are:

\begin{align}
  \label{eq: Q3oneacttrans}
  \tilde{q}_{12} &= \tilde{F}_{12} \frac{c}{\tilde{H}_{12} + c},\\
    \label{eq:q21CCOoneacttrans}
  \tilde{q}_{21}& =
          \frac{ \left( k_{13} c -
          {k_{23}} \right)^2}{ k_{13}
          k_{31} c +
          k_{23}
          k_{32}}
          \frac{k_{31}k_{32}}{k_{31} + k_{32}}  = \tilde{F}_{21} \frac{\left( c -
         \frac{k_{23}}{k_{13}}\right)^2}{\tilde{H}_{21} + c} 
  \\
  \label{eq:q23CCOoneacttrans}
  \tilde{q}_{23} & =
                             \frac {k_{13}    k_{31} c +    k_{23}  k_{32}}
                              {
                              k_{31} + k_{32}
                              }  = \tilde{k}_{23} c + \tilde{K}_{23}\\
  \label{eq:q32CCOoneacttrans}
   \tilde{q}_{32} &= k_{31} + k_{32}
\end{align}

with

\begin{align}
  \label{eq:F12}
  \tilde{F}_{12}  &= \frac{k_{23}(k_{31} + k_{32})}{k_{31}},
                    \tilde{H}_{12} = \frac{k_{23} k_{32}}{k_{13}
                    k_{31}}\\
  \tilde{F}_{21}  &=\frac{k_{13} k_{32}}{ k_{31} + k_{32}} ,
                    \tilde{H}_{21} = \tilde{H}_{12} = \frac{k_{23}
                    k_{32}}{k_{13} k_{31}}\\
    \tilde{k}_{23}  &=\frac{k_{13} k_{31}}{ k_{31} + k_{32}} ,
                    \tilde{K}_{23} = \frac{k_{23} k_{32}}{k_{31}  + k_{32}}
\end{align}

For a \rC\gO\rC\ model with one activating binding site and one
ligand-independent, we find that $\tilde{q}_{12}$ has the form of a
Michaelis-Menten term with maximum rate $\tilde{F}_{12}$ and
half-saturation constant~$\tilde{H}_{12}$. For~$\tilde{q}_{21}$ we
obtain another Hill function type term with a numerator of degree 2
and denominator of degree 1. This means that for large~$c$,
rate~$\tilde{q}_{21}$ behaves like a mass action term, although,
initially, it even decreases until vanishing at
$\frac{k_{23}}{k_{13}}$. The rate~$\tilde{q}_{23}$ is a mass action
rate boosted with a constant influx~$\tilde{K}_{23}$. Thus, for a
\rC\gO\rC\ model with one activating and one ligand-independent
transition to the open state, a qualitatively different model is
obtained. Rather than linearly increasing with the ligand
concentration~$c$, the transition
rate~$\tilde{q}_{12}$ 
shows a saturating behaviour---it tends to the maximum
rate~$\tilde{F}_{12}$ for large~$c$. The rate~$\tilde{q}_{21}$ is
again ligand-dependent with a more complicated dependency on the
ligand concentration~$c$ which, however, behaves like a mass action
term for large~$c$. The transition~$\tilde{q}_{23}$ is a mass action
term, but with an additive ligand-independent
constant~$\tilde{K}_{23}$.

\subsubsection{One activating and one inhibitory binding site}
\label{sec:Q3ligandactinh}

We now consider the case that one of the two binding site facilitates
the transition from~\rC[1] to the open state~\gO[3] whereas the other
one increases the rate to the closed state~\rC[2] i.e. the channel has
one activating and one inhibitory binding site

\begin{align}
  \label{eq:Q3oneactoneinh}
  q_{13} = k_{13} c,   q_{23} = k_{23},   q_{31} = k_{31},   q_{32} =
  k_{32} c.
\end{align}

Using \eqref{eq:q12CCO}-\eqref{eq:q32CCO}, the rates of the
model~\rC\rC\gO\ are:

\begin{align}
  \label{eq: Q3oneactoneinhtrans}
  \tilde{q}_{12} &= \frac{k_{13} k_{23} (k_{31} + k_{32}c) }{k_{13}
                   k_{31} +
                   k_{23} k_{32}} = \tilde{k}_{12} c + \tilde{K}_{12}, \\
    \label{eq:q21CCOoneactoneinhtrans}
  \tilde{q}_{21}& =
          \frac{ \left( k_{13} c -
          {k_{23}} \right)^2}{ k_{13}
          k_{31}  +
          k_{23}
          k_{32}}
          \frac{k_{31}k_{32}}{k_{31} + k_{32} c}  = \tilde{F}_{21} \frac{\left( c -
         \frac{k_{23}}{k_{13}}\right)^2}{\tilde{H}_{21} + c} 
  \\
  \label{eq:q23CCOoneactoneinhtrans}
  \tilde{q}_{23} & =\frac {k_{13}    k_{31} +    k_{23}  k_{32}}
                              {
                              k_{31} + k_{32} c
                              }  c=  \tilde{F}_{23} \frac{c}{\tilde{H}_{23} + c}\\
  \label{eq:q32CCOoneactoneinhtrans}
   \tilde{q}_{32} &= k_{31} + k_{32} c = \tilde{k}_{32} c + \tilde{K}_{32}
\end{align}

with

\begin{align}
  \label{eq:6}
  \tilde{k}_{12}  &= \frac{k_{12} k_{23}k_{32}}{k_{13}k_{31} +
                    k_{23}k_{32}}, 
                    \tilde{K}_{12} = \frac{k_{12} k_{23}k_{31}}{k_{13}k_{31} + k_{23}k_{32}},\\
  \tilde{F}_{21}  &=\frac{k_{13}^2 k_{31}}{ k_{13}k_{31} + k_{23} k_{32}} ,
                    \tilde{H}_{21}= \frac{k_{31}}{k_{32}}\\
    \tilde{F}_{23}  &=\frac{k_{13} k_{31} + k_{23} k_{32}}{k_{32}} ,
                    \tilde{H}_{23} =  \tilde{H}_{21}=
                      \frac{k_{31}}{k_{32}}, \\
  \tilde{k}_{32}&= k_{32}, \tilde{K}_{32}=k_{31}.
\end{align}

For a model with one activating and one inhibitory binding site, we
see similar terms as for the model with one activating and one
ligand-independent transition that we considered in the previous
section. Here, $q_{12}$ and~$q_{32}$ are mass action terms with an
additional ligand-independent offset. The transition to the open
state, $q_{23}$, is again represented by a Michaelis-Menten term
whereas the rate~$\tilde{q}_{21}$ tends to a mass action rate for
large~$c$.


\subsubsection{Summary}
\label{sec:summaryBinding}

Our analysis of different mechanisms of ligand binding in the
model~\rC\gO\rC\ has shown that by transformation to the model
\rC\rC\gO, the dependency of some rates on the ligand
concentration~$c$ might be changed from mass action kinetics to a
Michaelis-Menten term, see Sections~\ref{sec:Q3ligandoneact}
and~\ref{sec:Q3ligandactinh}.  In addition, even if all rates in the
transformed model~\rC\rC\gO\ are mass action terms, an additional
ligand dependency might appear (here, in the transition~$q_{21}$
between the closed states~\rC[2] and \rC[1], see
Section~\ref{sec:Q3ligandtwoact})
. It remains to be seen if these differences are sufficient
to allow, in practice, to distinguish between independent binding
sites (as represented in the model~\rC\gO\rC) and sequential binding
(as in the model~\rC\gO\rC) by comparing fits of both model structures
to experimental data.




\section{Discussion}
\label{sec:discussion}


We have given an overview of the identifiability theory of aggregated
Markov models. Aggregated Markov models appear naturally in the
modelling of ion channels which generate stochastic dynamics~$\dyn$
that consists of alternating sequences $\tO[i]$ and~$\tC[j]$ of open
and closed times. Using aggregated Markov models, this dynamics is
described by transitions between (usually multiple) open and closed
states~\gO[m] and \rC[n].

We have first explained how aggregated Markov models can be enumerated
i.e. we have shown how the number of different aggregated Markov
models for a given number $n=\nO+\nC$ of states can be calculated
using P\'{o}lya enumeration. The results, which can be found in
Table~\ref{tab:enumeration}, clearly show that the number of
aggregated Markov models grows extremely rapidly---for only~$\nV=10$
vertices, we have more than 10 billion models! If we are inclined to
interpret each model as a representation of a different biophysical
``process'', the results in Table~\ref{tab:enumeration} are somewhat
reduced due to non-identifiability but nevertheless increases quickly
with the number of vertices.

The phenomenon of non-identifiability appears somewhat
counter-intuitive at first glance---how can a ``biophysical process''
that is represented by transitions between the open and closed states
of a Markov model not be reconstructed from the sequence of open and
closed times generated by this ``process''? In order to gain a better
understanding of this problem, let us consider an open
time~$\tO[i]$. Due to the presence of multiple open states, it is not
clear, by which of the open states~\gO[n]\ the open time~$\tO[i]$ has
been generated---or if, during the time~$\tO[i]$, the model has even
alternated between multiple open states. This raises the question how
much detail of the parameters and/or the structure of an aggregated
Markov model can be inferred from the dynamics~$\dyn$ i.e. a sequence
of open and closed times at all.

To answer these questions, we have not related models directly to the
dynamics~$\dyn$. Instead, we have considered reparametrisations of the
infinitesimal generator~$Q$ of a given aggregated Markov model. This
relies on a classical result by \citet{Fre:85a, Fre:86a} that the
dynamics~$\dyn$ of an aggregated Markov model with~$\nO$ open
and~$\nC$ closed states is characterised by the bivariate
distributions~$\fOC(\tO,\tC)$ and~$\fCO(\tC,\tO)$. This implies that
reparametrising a model~$Q$ so that the distributions~$\fOC(\tO,\tC)$
and~$\fCO(\tC,\tO)$ remain unchanged will yield a model~$\tilde{Q}$
whose dynamics~$\dyn$ is indistinguishable from the original model~$Q$
i.e. $Q$ is \emph{not
  parameter-identifiable}.

If such a model~$Q$ which lacks parameter identifiability is used for
modelling ion channels, it will represent the dynamics $\dyn$ 
as well as any reparametrised version~$\tilde{Q}$ of $Q$. 
Thus, the choice of parameters~$q_{ij}$ for representing the
dynamics~$\dyn$ is not unique---multiple different parameter set (as
we have seen, in most cases, infinitely many) represent the dynamics
equally well.

If our primary aim is to find a model that reproduces the
dynamics~$\dyn$ accurately, this might not be considered a very
important problem. Indeed, phase type distributions
\citep{Neu:78a}---which can be interpreted as sojourn time
distributions~$\fC$ of an aggregated Markov model with only one open
state---are popular generalisations of the exponential distribution
for waiting times. Here, non-identifiability is not a concern because
the only role of the states (``phases'') is to accurately describe a
given data set, not to represent particular ``states'' of a
system. But if our goal is to build aggregated Markov model that
represent aspects of the underlying biophysics of the observed
dynamics, not being able to unambiguously determine the rates~$q_{ij}$
implies that a non-identifiable model does not allow us to gain
insight into the transition rates between different biophysical
states.

We have investigated the consequences of non-identifiability by
considering the fully connected three-state model. For a given
parameter set, there usually exists an infinite equivalence class of
models. Under certain conditions, this ambiguity could be resolved by
reducing the non-identifiable fully-connected model which depends on
six rates to either of the parameter-identifiable models~\rC\gO\rC\
or~\rC\rC\gO which depend on four rates. 

However, crucially, this model reduction does not allow us to
\emph{rule out} the presence of particular transitions between
conformational states that are represented in the fully connected
model. Rather, this only demonstrates that given dynamics~$\dyn$
\emph{can} be explained in the absence of some of these transitions
between states of the channel protein. But going further, model
reduction does not even allow us to distinguish between the
models~\rC\gO\rC\ and ~\rC\rC\gO---as we have seen, if a parameter set
of the fully connected three-state model can be reduced to the model
structure~\rC\gO\rC\ it can also be reduced to~~\rC\rC\gO\ and vice
versa. And despite the fact that the parameters~$q_{ij}$ are
identifiable for either of the two models, the \emph{model
  structure}~$\mathcal{G}$ for models with~$\nO=1$ open and~$\nC=2$
closed states remains non-identifiable---it is not possible to decide
if~\rC\gO\rC\ or~\rC\rC\gO\ provides a better representation of the
dynamics~$\dyn$. This shows that it is not possible to distinguish if
the dynamics~$\dyn$ is generated by one ``fast'' and one ``slow''
closed state from which the channel can transition to the open
state---this mechanism is represented by the model~\rC\gO\rC. Or,
alternatively, via a stepwise process where the channel makes a
transition from a closed state~\rC[1] to a state~\rC[2] which primes
the channel for entering the open state~\gO[3]. The two mechanisms
are, obviously, also very different from a biophysical point of
view---if we assume, as in Section~\ref{sec:biophysics}---that the
transitions between states are regulated by binding sites, the
model~\rC\gO\rC\ represents a channel with two independent binding
sites whereas~\rC\rC\gO\ is based on the assumption of sequential
binding.

Interestingly, reducing the non-identifiable fully connected
three-state model to an identifiable model with four instead of six
rates is not always possible because, for some parameter sets,
reparametrising the model would lead to negative rates. 
In this situation, the non-identifiability cannot be resolved by
choosing a particular model, simply because it is not obvious which
model to pick as a representative for the infinite sets of equivalent
models of the equivalence class. Interpreted from a biophysical point
of view, in this case it is not possible to infer, which
conformational changes might have generated the observed
dynamics~$\dyn$. In the case of the fully connected three-state
models, this applied to models for which the sojourn time
distribution~$\fC$ could not be represented as a mixture of
exponential distributions but as a signed mixture of
exponentials. This shows that some dynamics~$\dyn$ that aggregated
Markov models can 
generate, cannot be represented with an identifiable model, or, in
other words, some dynamics~$\dyn$, in principle, cannot be associated
unambiguously with a biophysical mechanism.


{In summary, two lessons can be learnt.} First, for some parameter sets,
the fully connected model can be reduced to the model
structures~\rC\gO\rC\ and~\rC\rC\gO. This means that the dynamics
\emph{can} be represented by assuming a simpler, acyclic
mechanism. But this does not exclude that the dynamics could equally
well be generated by a mechanism where transitions between all states
are possible. Also, it is not possible to distinguish between the
fundamentally different mechanisms~\rC\gO\rC\ and~\rC\rC\gO. Second,
the fully connected three-state model can generate sojourn time
distributions based on a signed mixture of exponential whose densities
show local maxima and minima rather than the monotonously decreasing
densities obtained from a ``normal'' mixture of exponential
distributions (where all coefficients of the exponentials are positive
and form a finite probability distribution). These sojourn time
distributions cannot be represented so that the model is
parameter-identifiable so that these dynamics cannot be associated
with transitions between different conformational states of the
channel protein.

It has been suggested that the problem of identifiability of model
structure (i.e. of the underlying biophysical mechanism) might be
addressed by considering multiple data sets. 
We have investigated the question if the two alternative three-state
models~\rC\gO\rC\ and~\rC\rC\gO\ could be distinguished when the
dependency of a channel on the concentration~$c$ of a ligand is
considered. Starting from models based on particular assumptions for
different ligand binding sites, we have investigated how the ligand
dependencies differ in the transformed model~\rC\rC\gO\ after
reparametrising the model~\rC\gO\rC. For some models, our results show
that some transitions in the transformed model~\rC\rC\gO\ are enzyme
kinetics whereas in the original model~\rC\gO\rC, mass action was
assumed. This does not help to distinguish~\rC\gO\rC\ and~\rC\rC\gO\
because alongside mass action kinetics, Michaelis-Menten terms (or
other Hill functions) are common models of ligand binding in ion
channel modelling. A clearer hint that dynamics~$\dyn$ could be better
described by the \rC\gO\rC\ model than the~\rC\rC\gO\ model, might be
the emergence of ligand-dependent transitions where they would
normally not be expected---the most striking example is the case of
the~\rC\gO\rC\ model with two activating binding sites, see
Section~\ref{sec:Q3ligandtwoact}: here, the mass action
rates~$q_{13}(c)$ and~$q_{23}(c)$ connecting the closed states~\rC[1]
and~\rC[2] to the open state~\gO[3] in the model~\rC\gO\rC\ translate
to analogous rates~$q_{12}(c)$ and~$q_{23}(c)$ in the transition from
closed state~\rC[1] via~\rC[2] to the open state~\gO[3] in the
model~\rC\rC\gO. But the emergence of a mass action rate~$q_{21}(c)$
in the transition from~\rC[2] to~\rC[1] is unexpected and would
unlikely be chosen as such if the model had been developed based on
mechanistic assumptions
. It remains to be
seen if criteria like this can be exploited in practice when comparing
the fits of equivalent ion channel models to the same data set.

Having explored the non-identifiability of aggregated Markov models, we
will reflect on implications for modelling ion channels and indicate
how mechanistic data-driven models can nevertheless be developed under
these circumstances. 
Beyond the sufficient condition of $2\nO \nC$ for the maximum number
of rate constants of an aggregated Markov model introduced by
\citet{Fre:85a, Fre:86a}, to the best of our knowledge, no other
criteria are available that can be applied easily to decide if a given
Markov model is identifiable. If it is unknown if a given model is
identifiable, indirect information can be gained by using approaches
that investigate the uncertainties of parameters. We have suggested
specifically that Markov chain Monte Carlo (MCMC) can be useful for
deciding if a model could be non-identifiable
\citep{Sie:12b}. Alternatively, profile likelihood approaches have
been applied to a wide range of models to assess identifiability
\citep{Bat:88a, Rau:09a, Kre:12a, Kre:13a, Sim:23a, Cio:24a, Pla:24a}
although, to the best of our knowledge, they have not been applied
specifically to aggregated Markov models.

An approach that attempts to address both the lack of parameter
identifiability as well as non-identifiability of model structure is
the use of \emph{canonical forms}. Canonical forms as those proposed
by \citet{Bau:87a,Kie:89a} and \citet{Lar:98a, Bru:05a,Flo:06a} are model structures that
are on the one hand parameter identifiable and on the other hand,
provide representatives for each of the different model structures
that exist for a given number of open and closed states. Restricting
the set of models to be considered to representatives of an
equivalence classes of models makes the process of model selection
more efficient because it prevents the modeller from comparing fits of
equivalent models (such as the models~\rC\rC\gO\ and~\rC\gO\rC\
investigated in detail in this study) which, in theory, should show
identical performance when fitted to the same data. Although canonical
forms solve the statistical problems related to non-identifiability
when it comes to fitting aggregated Markov models to data, modellers
are restricted the model structures defined by the canonical forms
which constrains their freedom of representing biophysical
mechanisms. A more severe problem is that some data sets might only be
represented by physically unrealistic models, for example, the
\emph{manifest interconductance rank (MIR)} form proposed by
\citet{Bru:05a} as well as the BKU form \citep{Bau:87a,Bru:05a} might
only be able to model some data sets if some rates are allowed to be
negative, similar to the example presented in
Section~\ref{sec:nonidExample}
. Nevertheless, the MIR form by \citet{Bru:05a} allows for a refined
representation of the equivalence classes of aggregated Markov models
because it considers the ``interconductance rank''~$p_{\text{\gO\rC}}$
i.e. the rank of the transition matrices~$Q_{\text{\rC\gO}}$
and~$Q_{\text{\gO\rC}}$ which consist of the rates connecting open and
closed states.  In MIR form, the number of transitions between open
and closed states is chosen to coincide with the rank
of~$Q_{\text{\rC\gO}}$ and~$Q_{\text{\gO\rC}}$. The MIR form therefore
depends on~$2 p_{\text{\gO\rC}} (\nO+\nC-p_{\text{\gO\rC}})$
parameters, consistent with the refined upper bound for the maximum
number of parameters derived in \citet{Fre:86a}.

It remains a challenging problem to design 
models that represent the observed dynamics~$\dyn$ of an ion channel
but also capture biophysical processes such as conformational
changes. Designing models based on detailed knowledge of activating
and inhibitory binding sites and other molecular properties of the ion
channel leads to models with complex structures consisting of a large
number of states and many parameters. Thus, this approach is fraught
with problems related to non-identifiability, or, even more simply,
over-parametrisation. 
Building models that mimics aspects of the molecular structure of ion
channels is based on the hope that the parameters of these models 
will reflect the
transition rates between biophysical states of the ion channel and
thus provide insights into the time scales of conformational changes.

We would like to contrast this with alternative approaches that
provide more direct insight into the time scales of conformational
changes by rigorous statistical analysis of the dynamics~$\dyn$. Modal
gating is a phenomenon that describes ion channels switching between
different dynamical patterns characterised by different levels of
activity (i.e. for example different levels of open
probability). These modes form a limited repertoire of dynamical
patterns between which the ion channel alternates
instantaneously. Thus, from a mathematical point of view, modal gating
is an example of dynamics that exhibits two clearly different time
scales. The ``fast'' dynamics is characterised by different modes that
are associated with particular dynamical patterns. On the ``slow''
time scale the ion channel switches between these modes.

Modal gating has been investigated statistically by
\citet{Ion:07a,Sie:14a}. Both have developed methods which enables us
to quantitatively infer the time scale of switching between modes as
well as characterise the dynamics of individual modes. It is
well-established that modes are manifestations of different
conformational states in the dynamics of an ion channel, see, in
particular, the study \citet{Cha:11a} as well as a wealth of
additional references reviewed in the Discussion of
\citet{Sie:14a}. Thus, statistical analysis of modal gating enables us
to infer the dynamics of conformational changes of the ion channel by
associating modes with different conformational states of the ion
channel. The importance of modal gating for the modelling of ion
channels is increasingly recognised---modal gating dynamics was
included in~\citet{Ull:12a} as one of multiple data
sources. \citet{Sie:12a} and \citet{Bic:16a} built models whose
structures were designed based on modal gating as the underlying
construction principle and \citet{Sie:16a} proposed a structure
specifically for representing modal gating as a hierarchical
process---the hierarchical Markov model combines a Markov model that
represents the switching between different modes with models for the
dynamics of the individual modes to a model that accounts for
both switching between modes as well as the dynamics within modes. Due
to the close relationships of modes and conformational states, the
states of the hierarchical Markov model can be interpreted to
represent conformational changes, similar to
models 
based on certain assumptions on biophysical processes such as ligand
binding. But in contrast to these models, in the hierarchical Markov
model, the rates between different conformational states can be
parameterised based on a rigorous statistical analysis of the
transitions between modes observed in the data.

Additional insight into both the dynamics as well as the molecular
structure of ion channels can be gained from data that characterises
the latency of an ion channel i.e.  the delayed response of an ion
channel to changes in ligand concentrations. \citet{Haw:25a} developed
a method for extending an existing model whose ligand dependency was
parametrised using multiple data sets at constant concentrations~$c$
by integrating additional data that investigated the response of the
channel to changes of the ligand concentration. The biophysical
significance of this model can best be understood by considering the
transition from a ligand concentration~$c_1$ to a ligand
concentration~$c_2$ where for concentration~$c_1$ the channel protein
is in conformation A with a high probability whereas for
concentration~$c_2$ the channel is most likely in different
conformational state~B. When instantaneously switching from~$c_1$
to~$c_2$ the channel cannot immediately adjust to the new ligand
concentration, instead it transitions from A to B with a delay. This
delay is represented in the model by introducing an integral term that
averages the ligand concentration over a certain finite time interval.
Thus, instead of an immediate switch the integral introduces a gradual
transition from~$c_1$ to~$c_2$ and instead of an immediate ``jump''
the transition from conformation A to conformation B comes with a
delay.

In summary, our study has demonstrated that non-identifiability poses
difficult challenges for developing models that represent both the
dynamics~$\dyn$ as well as the biophysics of an ion channel by
identifying individual states of a Markov model with conformational
states of the channel protein. But we have presented some alternative
approaches for developing 
models that represent conformational changes underlying the opening
and closing of an ion channel by integrating additional data. By
giving this detailed introduction into the complex phenomenon of
non-identifiability we hope to facilitate the task of building
mechanistic data-driven models of ion channels and, more general,
taking advantage of the framework of aggregated Markov models for
other applications.

\backmatter





\bmhead{Acknowledgements} {I would like to thank an
  anonymous reviewer for their supportive and helpful comments.} I
would {also} like to thank J\"urgen Will for generously
sharing his knowledge of P\'olya enumeration 
and, in particular, for explaining that graphs with two colours can be
enumerated by representing them as rooted graphs. Also, I am very
grateful for generous travel funding provided by the MATRIX research
institute (Creswick, VIC, Australia) via a MATRIX-Simons grant for
providing generous funding. In particular, I would like to thank
Matthew Simpson, Ruth Baker, Oliver Maclaren and Jennifer Flegg for
organising the excellent workshop ``Parameter Identifiability in
Mathematical Biology'' where parts of this work were presented.

\bmhead{Data availability} This study has no associated data.

\bibliography{/Users/merlin/Documents/references/refer}

\end{document}